%% file: script_Materealistic.tex
\documentclass[12pt,letterpaper]{article}
\usepackage[a4paper, total={7in, 10in}]{geometry}

\usepackage{graphicx}
\usepackage{rotating}
\usepackage{multirow}
\usepackage{helvet}
\usepackage{authblk}
\usepackage{hyperref}
\usepackage{amsmath} 
\usepackage{amssymb} 
\usepackage{orcidlink} 
\usepackage[super,comma,sort&compress]  
   {natbib}
\usepackage{textcomp}
\usepackage{siunitx}
\usepackage{booktabs}
\usepackage{subcaption}
\usepackage{graphicx}

\usepackage{setspace}
\makeatletter
\renewcommand{\maketitle}{\bgroup\setlength{\parindent}{0pt}
\begin{flushleft}
  \textbf{\@title}
  \@author
\end{flushleft}\egroup}
\makeatother

\title{Materealistic? How European energy system models exceed raw material reserves\\[1em]}
\date{}

\author[1,3,*,\orcidlink{0000-0002-8512-4541}]{Jan Mutke}
\author[2,\orcidlink{0000-0001-7170-103X}]{Jonas Finke}
\author[2, \orcidlink{0009-0005-7867-5934}]{Katharina Esser}
\author[1,3, \orcidlink{0000-0001-9536-9465}]{Heidi Heinrichs} 

\affil[1]{Forschungszentrum Jülich GmbH, Institute of Climate and Energy Systems, Jülich Systems Analysis, Wilhelm-Johnen-Straße, 52428 Jülich, Germany}
\affil[2]{Ruhr University Bochum, Universitätsstr.~150, 44801 Bochum, Germany}
\affil[3]{University of Siegen, Chair for Energy Systems Analysis, Department of Mechanical Engineering, 57076 Siegen, Germany}

\affil[*]{Correspondence: j.mutke@fz-juelich.de}

\begin{document}

\maketitle
 
\section*{SUMMARY}

Decarbonising energy systems reduces emissions and fossil fuel dependency, but expanding renewables increases demands for critical raw materials.
Most energy system models, however, neglect material demands, putting the material feasibility of energy scenarios at question.
We combine a systematic review of 59 highly decarbonised European energy system modelling studies with a quantitative ex-post assessment of material demands for 5 key technologies and 19 materials.
We find that material demands exceed Europe's population-based shares of current global reserves for seven materials (Ga, In, Ir, Te; less pronounced for Ag, Se, V), in particular if multiple sectors of the energy system are considered.
Competing non-energy demand further amplifies the scarcity, while technological innovation can either alleviate or intensify it.
We conclude that energy efficiency, recycling, expanding reserves and technological innovation may only partly address the identified shortages and call for energy sufficiency measures to achieve sustainability in the energy-material nexus.


\section*{INTRODUCTION}

With the transition from fossil to renewable energy sources, greenhouse gas emissions and the consumption of fossil fuels decrease.
This transition is necessary to mitigate the negative impacts of climate change and reduce fuel import dependencies.
But the expansion of new energy technologies, in particular wind turbines, photovoltaics (PV), battery storages and electrolysis and concentrated solar power (CSP), leads to new and increased material demands \citep{GlobalCriticalMinerals2025} that must be navigated \citep{schulzeFossilFuelsMetals2024}.
Material supply, however, is limited by the concentrated distribution of resources and related geopolitical constraints, restricting the technical feasibility of expanding new energy technologies \citep{RegulationEU2024}.
The extraction and processing of many materials is moreover associated with negative impacts, e.g.\ environmental and social impacts \citep{berthetAssessingSocialEnvironmental2024}, and supply risk \citep{diasSupplyChainAnalysis}.
Considering materials in energy scenarios is therefore crucial to ensure feasibility  and reliability of the energy transition.
This has been recognised, for instance, by the International Energy Agency \citep{RoleCriticalMinerals}, the European Union \citep{europeancommissionStudyCriticalRaw2023}, and governments \citep{DOE_Crit_Ass_23, muddUK2024Criticality}.

Long-term energy scenarios have been generated and supported by energy system models (ESMs) for decades in research, policy making and industry \citep{pfenningerEnergySystemsModeling2014}.
These models typically find the least-cost system configuration, in particular installed capacities of renewable energy technologies, storages and sometimes electrolysers, that meet future energy demands.
In capacity optimisation, however, only few ESM studies consider material demands endogenously.
\citet{vaiMayAvailabilityCritical2025} study a decarbonisation pathway for the Italian energy sector until 2050 using TEMOA-Italy. They analyse how supply disruption scenarios for six materials, implemented through model constraints on the material availability, affect energy security.
Other studies consider impacts of material demands rather than a constrained material availability.
\citet{colucciCombinedAssessmentMaterial2025} analyse trade-offs between energy and material supply risks in a decarbonised Italian power sector. They consider material supply risk as an objective function in a multi-objective optimisation, based on current material supply chains.
\citet{raunerHolisticEnergySystem2017} consider metal depletion as one of 17 impact categories of a life cycle assessment (LCA) in a German power system study through multi-objective optimisation.
Finally, \citet{tokimatsuEnergyModelingApproach2017, tokimatsuEnergyModelingApproach2018} couple an ESM and a mineral resource balance model to study material demands of global net-zero scenarios.
All these studies, however, do not relate the material demands to available reserves.
Moreover, most ESMs, like the widely used open-source frameworks PyPSA \citep{brownPyPSAPythonPower2018}, Calliope \citep{pfenningerCalliopeMultiscaleEnergy2018} or OSeMOSYS \citep{howellsOSeMOSYSOpenSource2011} frameworks, do not consider raw material demands endogenously at all.
Therefore it remains unclear whether the modelled energy futures' material demand can be met, putting the feasibility and significance of the model outcomes at question.

Instead of the rare endogenous consideration of material demands, some studies assess these ESM outcomes ex-post.
For instance, \citet{martinEnergyFutureClimate2023} analyse material supply risks of European energy futures obtained with the Calliope model for 2030 and 2050.
\citet{junneEnvironmentalSustainabilityAssessment2020} and \citet{Fuss2021} analyse the use of minerals and metals as one impact category in an LCA of ESM-based scenarios.
Moreover, several studies assess material demands of energy scenarios not clearly originating from ESMs, but for instance from integrated assessment models (IAMs).
\citet{moreauEnoughMetalsResource2019} evaluate the metal demand of six global energy scenarios, among others from the IPCC, IEA and IRENA, for the year 2050.
\citet{schlichenmaierMayMaterialBottlenecks2022} study material bottlenecks in a global 1.5\textdegree~scenario from the European Commission.
\citet{Koljonen2024} evaluate material demands of global energy futures in line with the Nationally Determined Contributions based on modelling with the TIMES-VTT IAM.
These ex-post assessments, however, are only carried out for a limited amount of selected models or scenarios.
Therefore, the material demand and thus the feasibility of the vast majority of long-term ESM studies is still an open question. 

This literature gap calls for a systematic analysis of the material demand in the existing ESM literature.
Existing reviews, for instance by \citet{schulzeOvercomingChallengesAssessing2024}, analyse how material demands of future energy systems are considered in ESMs and IAMs.
In line with our above analysis, they found 63 studies performing ex-post analyses and only 9 studies in which materials were considered ``model-based''.
The latter include 4 studies with soft-linking and 5 with hard-linking approaches, a majority of which does not use ESMs.
\citet{schulzeOvercomingChallengesAssessing2024}, however, focus on the modelling methodology, and do not analyse the material demand themselves.
Moreover, \citet{montanaAssessingCriticalRaw2024} review critical raw materials in energy technologies more broadly.
They discuss five long-term scenarios, but also focus on the methodology used, not the resulting material demand itself.

In this paper, we therefore combine a systematic literature review of highly decarbonised European ESM studies with a quantitative ex-post material assessment.
First, we extract installed capacities of key future renewable technologies, namely PV, wind, electrolysers, batteries and CSP from 59 highly decarbonised European ESM studies.
Second, we compute the resulting demand of 19 selected raw materials that are required for the 5 key technologies and for which data are available.
Third, we relate these material demands to current global reserves to analyse the material-related feasibility of all energy futures modelled, similar to the analysis of a single energy scenario by \citet{schlichenmaierMayMaterialBottlenecks2022}. 
Throughout our analysis, we follow a lower boundary approach, i.e.\ we conservatively estimate material demands.
This allows us to identify which material demands exceed reserves with high fidelity without claiming that other materials are abundant.
Finally, we vary several of our assumptions regarding reserve allocation, competing non-energy material demands, sub-technology roadmaps and material intensities to analyse the robustness and drivers of our findings.
In particular, we answer the following research questions:
\begin{itemize}
    \item Do material demands of highly decarbonised European energy futures represented in energy system modelling studies exceed current reserves?
    \item What are drivers of potential excess demand and what strategies could be used to address them?
\end{itemize}

\section*{RESULTS}

\subsection*{Capacity expansion varies strongly across highly decarbonised European energy system models}

Expansion capacities could be extracted for 59 out of 142 eligible studies as shown in Figure \ref{fig:capacity}.
The extracted data for all 59 studies is shown Table \ref{tab:capacity} in the Appendix. 
Despite the broadly similar European scope and high level of decarbonisation, the expanded capacity varies strongly across studies, ranging from \SIrange{0.12}{5.1}{\tera\watt} for PV, \SIrange{0.69}{3.0}{\tera\watt} for onshore wind, \SIrange{0}{0.87}{\tera\watt} for offshore wind, \SIrange{0}{2.2}{\tera\watt} for electrolyser,  \SIrange{0}{6.2}{TWh} for batteries and \SIrange{0}{0.20}{\tera\watt} for CSP.
Sectoral coverage (electricity only versus multiple sectors) strongly correlates with the total capacity.
Splitting the 59 studies into the 29 studies with highest and 30 studies with lowest capacity, only 4 of the studies with highest capacity cover exclusively the electricity sector.
Conversely, only 9 of the studies with lowest capacity cover multiple sectors.
This correlation can be explained by an overall higher energy demand that must be met in multi-sector studies.
In contrast, the distinction between studies with full decarbonisation or net-zero systems and studies with lower (yet still high) degrees of decarbonisation has little explanatory power.
Differing capacities of further technologies, e.g.\ nuclear power, may partly explain the observed broad variations of renewable and storage capacities.
But following a lower boundary approach, we focus on those technologies that are main drivers of material demand, i.e. PV, wind, eletrolyser, battery and CSP.
Capacities of other technologies are not shown and analysed.

\begin{figure}[htpb]
    \centering
    \includegraphics[width=1.0\linewidth]{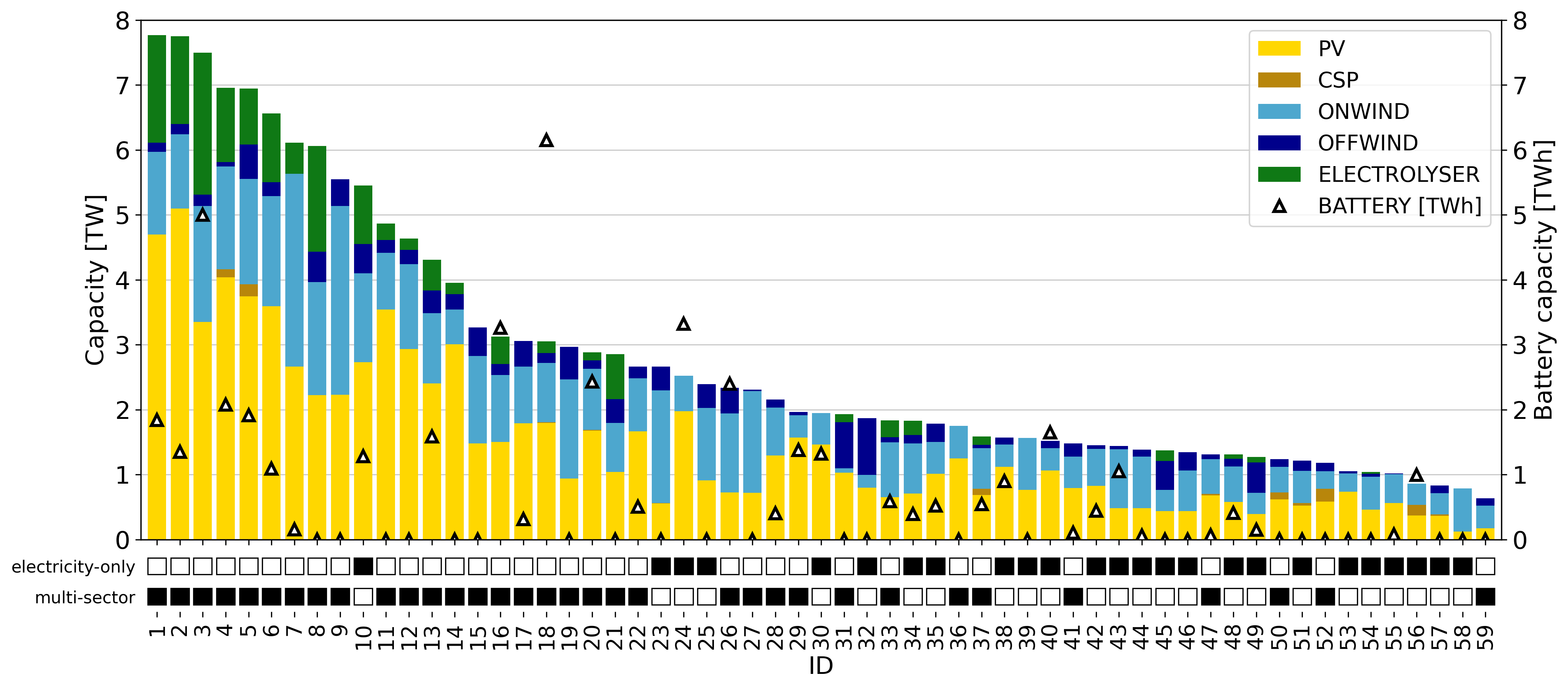}
    \caption{Capacity expansion of six energy technologies across 59 highly decarbonised European energy system modelling studies. The lower part additionally indicates the studies' sectoral coverage and level of decarbonisation.}
    \label{fig:capacity}
\end{figure}

\subsection*{Material demands for capacity expansion exceed reserves}

\begin{figure}[htpb]
    \centering
    \includegraphics[width=1.0\linewidth]{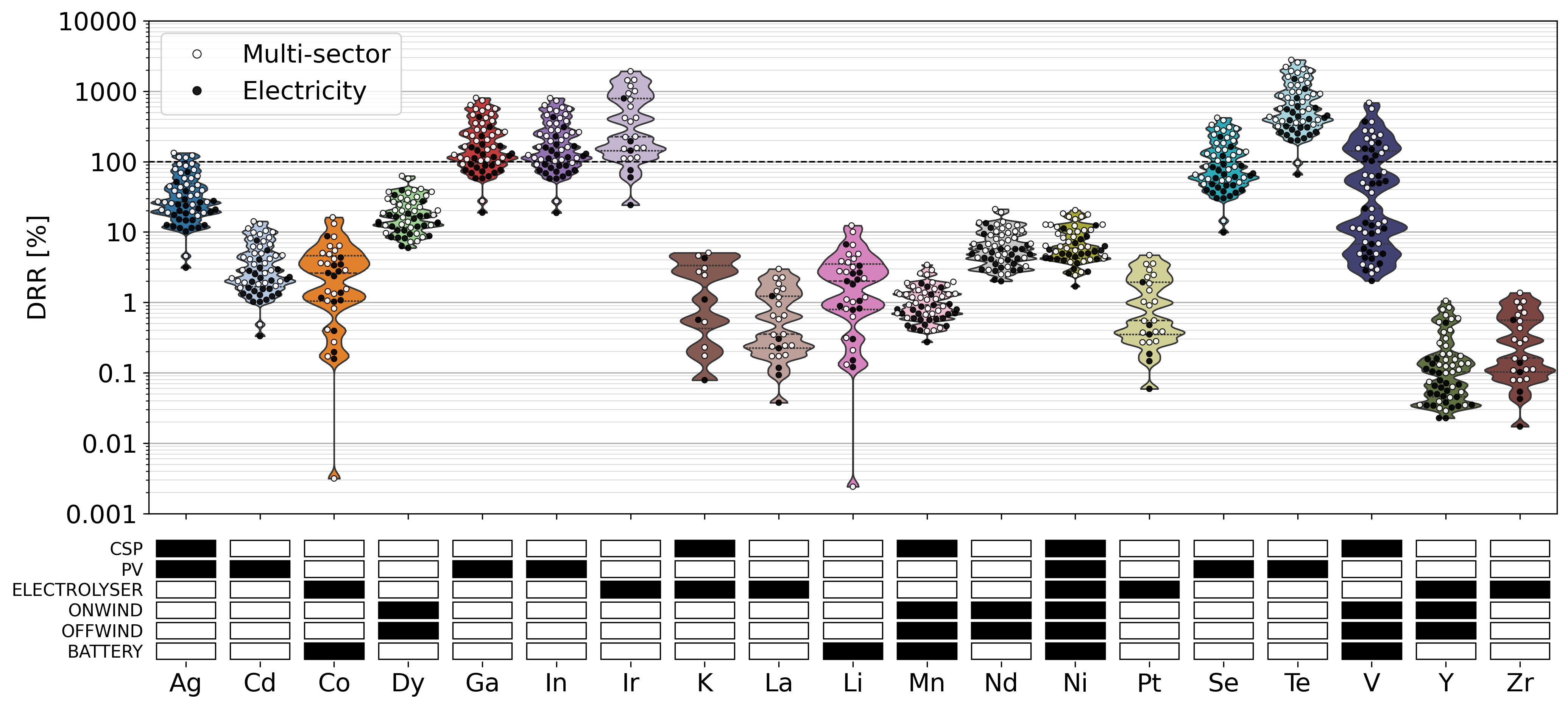}
    \caption{Distribution of demand-to-reserve ratio (DRR) across all studies. Above 100\% indicate that material demand exceeds the reserves allocated to Europe. Note the logarithmic scale. Coloured dots indicate whether a study is electricity only or multi-sectoral. The lower part additionally indicates in what technologies the respective materials are used.}
    \label{fig:reference}
\end{figure}

Based on the extracted capacities and technology-specific material demands, we obtain the total demand of 19 raw materials\footnote{Ag -- Silver, Cd -- Cadmium, Co -- Cobalt, Dy -- Dysprosium, Ga -- Gallium, In -- Indium, Ir -- Iridium, K -- Potassium, La -- Lanthanum, Li -- Lithium, Mn -- Manganese, Nd -- Neodymium, Ni -- Nickel, Pt -- Platinum, Se -- Selenium, Te -- Tellurium, V -- Vanadium, Y -- Yttrium, Zr -- Zirconium} for each of the 59 studies.
We relate this material demand to the current reserves, where we assume that a share of the global reserves is available to the European system in proportion to its population share (5.6\%, cf.\ \hyperref[sec:METHODS]{Methods}), i.e.\ a globally uniform per capita distribution.
The distribution of this \emph{demand-to-reserve ratio} (DRR) across all studies is shown in Figure \ref{fig:reference} for the \emph{reference case}.
In addition to a population-based reserve allocation, the reference case does neglects non-energy competing material demand, considers a conservative sub-technology roadmap, and assumes current material efficiencies.
The detailed definition of the reference case and the additional sensitivity cases can be found in the \hyperref[sec:METHODS]{Methods}.
For 7 out of 19 materials, studies with DRR above 100\% exist, i.e.\ the material demand exceeds the reserves allocated to Europe.
The demand of Ga, In, Ir and Te exceeds reserves by a factor of around 10 in the majority of studies and has a DRR median value significant above 100\%.
For Ag, Se and V, only a minority of studies exceeds the reserves.
Due to the lower boundary approach, we cannot claim that the demand for the remaining 12 materials with DRR below 100\% across all studies remains below the allocated reserves.
The findings for Ir and Te are robust against a change in global reserve allocation from a population-based to a GDP-based approach (5.6\% to 16.7\% of global reserves, see Figure \ref{fig:gdp_allocation} in the Appendix).
For Ag, Se and V, in contrast, no or only few studies show DRRs above 100\% for GDP-based allocation.
Ga and In show a mixed picture with 43 studies above and 16 below 100\% for this allocation approach.

\subsection*{Sectoral model coverage drives material demand}

Multi-sectoral coverage correlates with higher material demand for several materials, in particular for those materials used in wind and PV (cf.\ coloured dots and lower part in Figure \ref{fig:reference}).
These materials show similar distribution patterns.
Specifically for Ga, In, Ir and Se, this implies that most studies with DRRs below 100\% originate from ESM studies covering only the electricity sector.
Thus, almost all studies with more comprehensive energy sector coverage have demands above the reserves for these materials.
Some materials (Ir, K, La, Pt) are only used in selected technologies, e.g.\ electrolysers or CSP, that are not expanded or not expandable in all ESM studies.
Therefore, they show overall fewer dots and different distribution patterns.
Vanadium, notable by being used for many technologies through steel, exhibits no clear correlation between DRR and sectoral coverage.

\subsection*{Competing non-energy material demands increase demand-to-reserve ratios}

\begin{figure}[htpb]
    \centering
    \begin{subfigure}{0.57\linewidth}
        \centering
        \includegraphics[width=\linewidth]{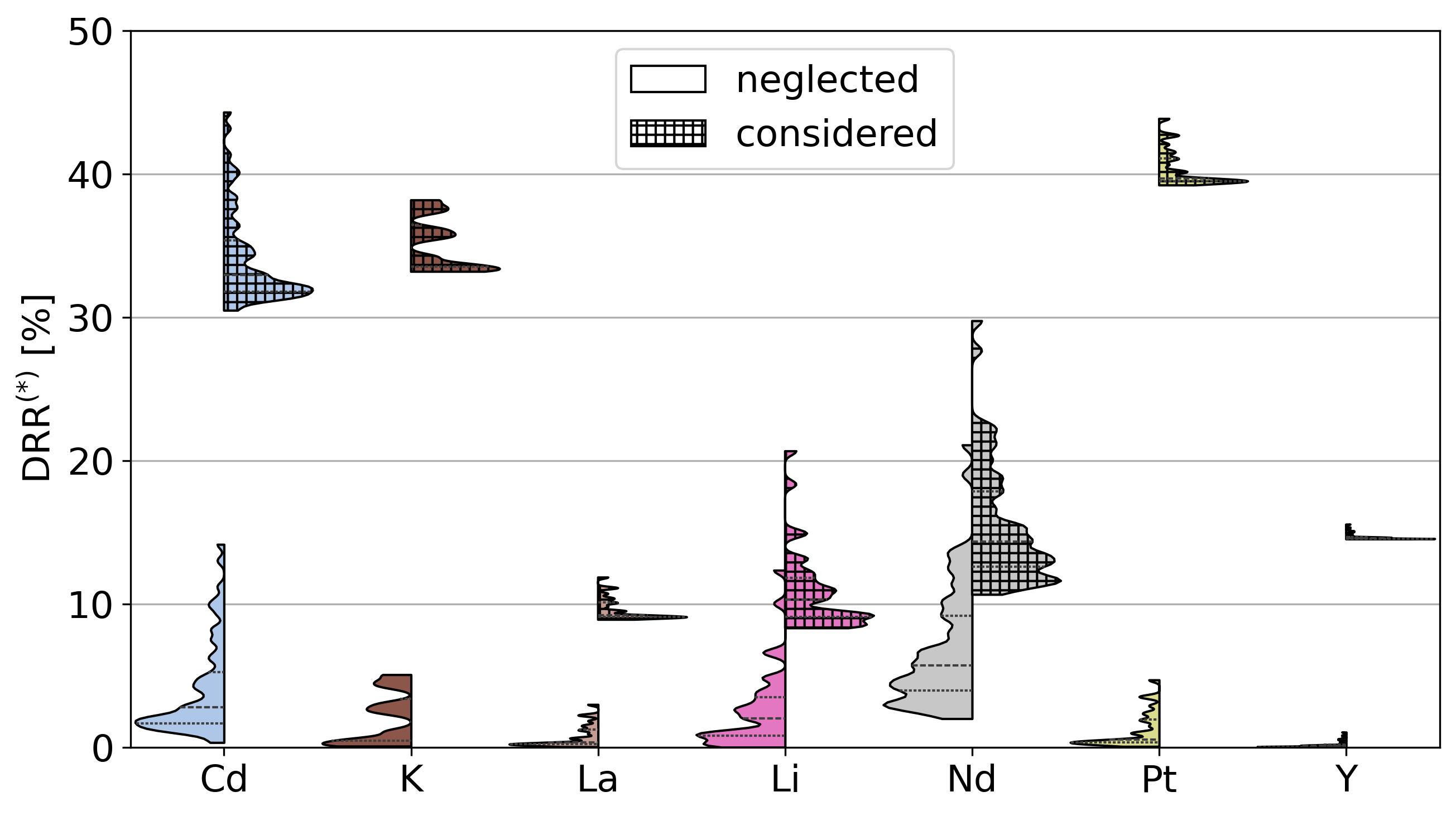}
        \caption{DRRs up to 50\%.}
    \end{subfigure}
    \hfill
    \begin{subfigure}{0.41\linewidth}
        \centering
        \includegraphics[width=\linewidth]{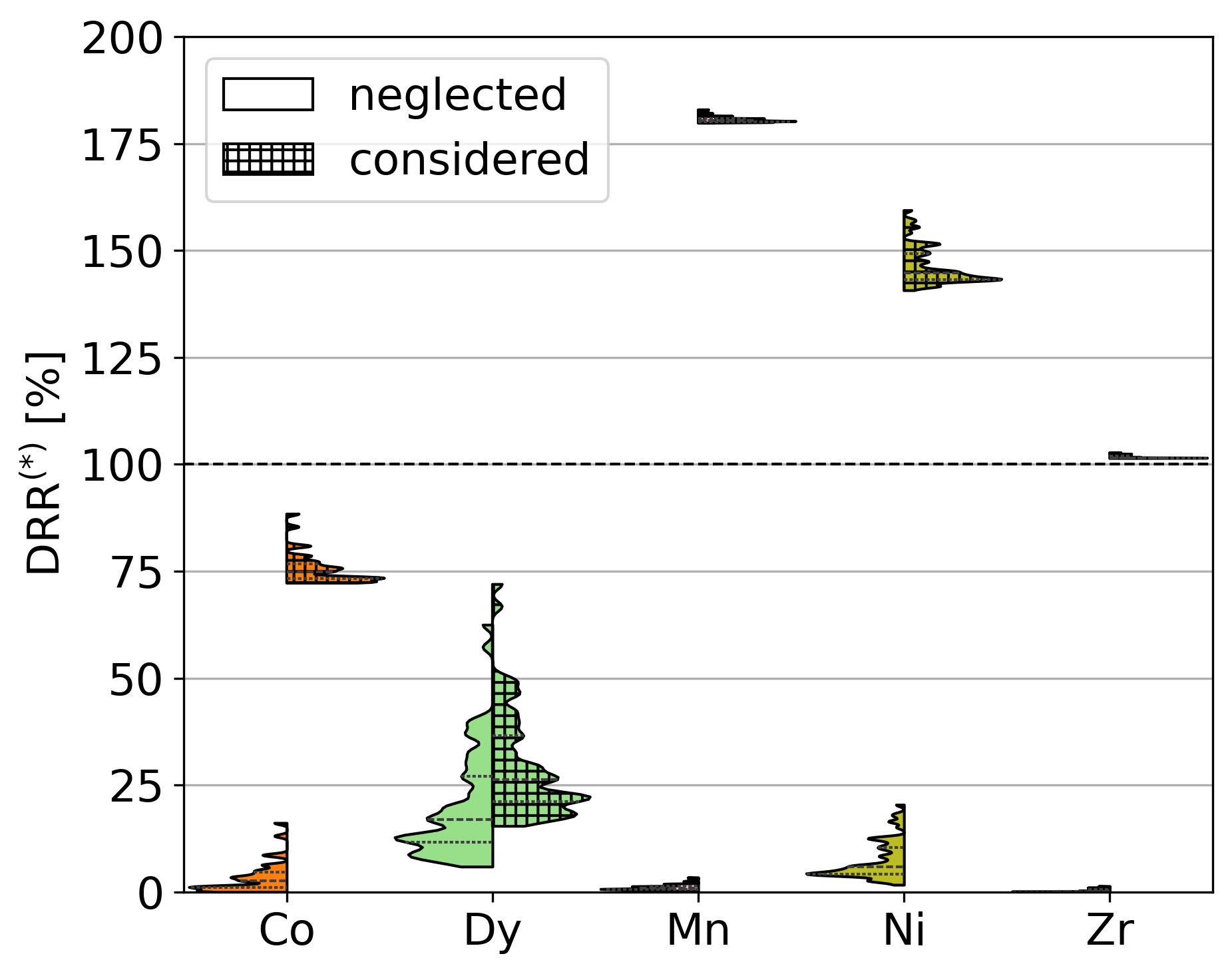}
        \caption{DRRs up to 200\%.}
    \end{subfigure}
     \hfill
    \begin{subfigure}{0.57\linewidth}
        \centering
        \includegraphics[width=\linewidth]{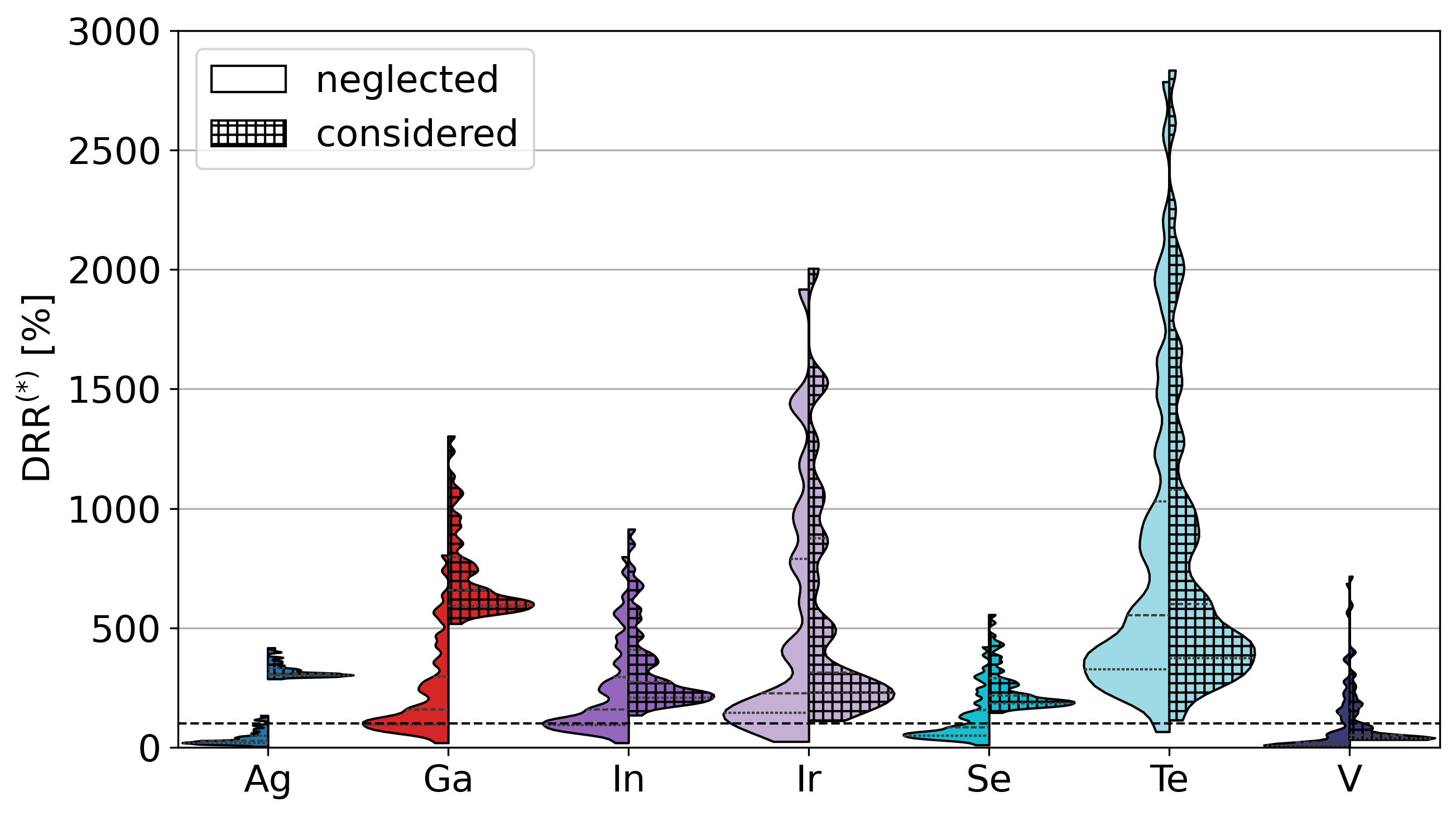}
        \caption{DRRs up to 3000\%.}
    \end{subfigure}
    \caption{Distribution of demand-to-reserve ratio (DRR) across all studies without ("neglected") and with competing ("considered") non-energy material demand added. Reserve allocation is population-based. Note the different value ranges and linear scale of the y-axes.}
    \label{fig:competing_demand}
\end{figure}

Figure \ref{fig:competing_demand} compares DRRs with and without competing non-energy material demand derived from \citet{schlichenmaierMayMaterialBottlenecks2022}.
Several observations can be made.
First, for the seven materials (Ag, Ga, In, Ir, Se, Te, V) with DRRs above 100\% in the reference case, the exceeding demand increases even further.
The difference is strongest for Ag and Ga, which have significant non-energy demand.
Second, Co, Mn, Ni and Zr have demands well below the reserves in the reference case, but DRRs around or above 100\% when considering competing demands.
For these materials, total demands may exceed allocated reserves, but not mainly driven by the considered energy technologies.
Finally, seven materials (Cd, K, La, Li, Nd, Pt, Y) reveal DRRs well below 100\% in both cases, i.e.\ the material demand from neither the added non-energy uses, nor from the considered energy technologies exceed allocated reserves.

\subsection*{Technological innovation may drive material demands -- in either direction}

\begin{figure}[htb]
    \begin{subfigure}{\linewidth}
        \centering
        \includegraphics[width=1.0\linewidth]{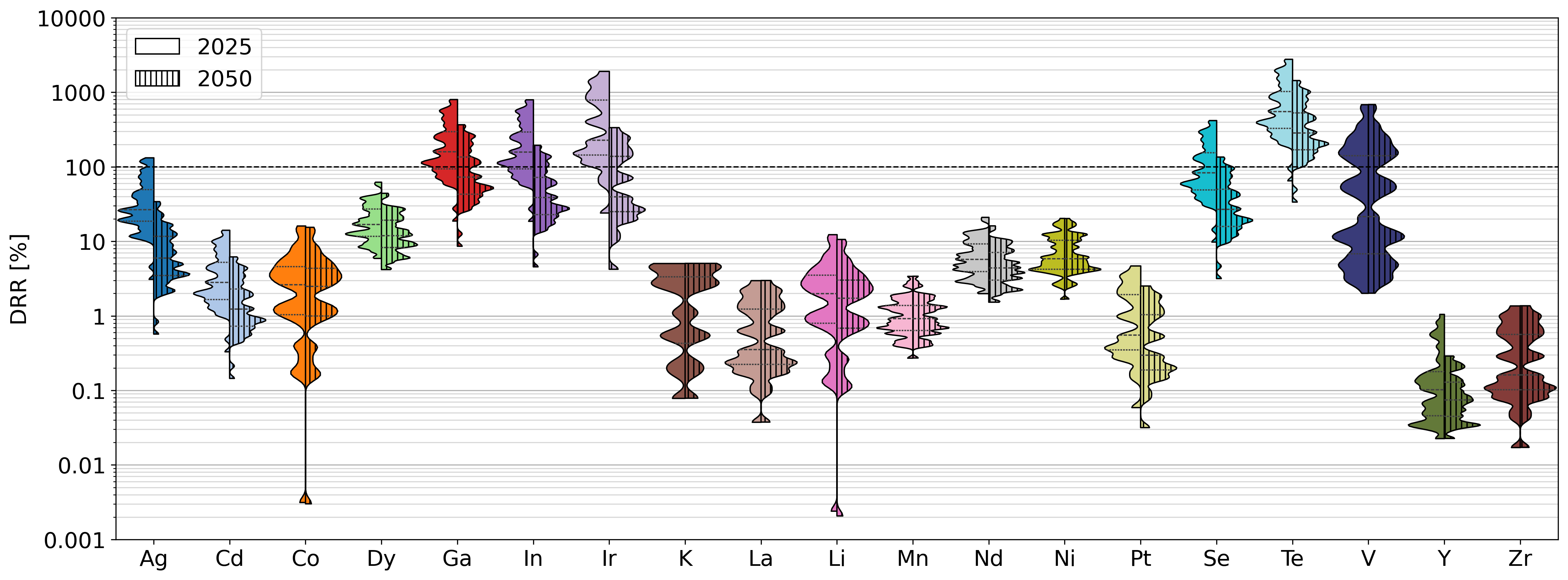}
        \caption{Current (left, year 2025) vs projected future (right, year 2050) material intensities.}
        \label{fig:material_intensity}
    \end{subfigure}
    \begin{subfigure}{\linewidth}
        \centering
        \includegraphics[width=1.0\linewidth]{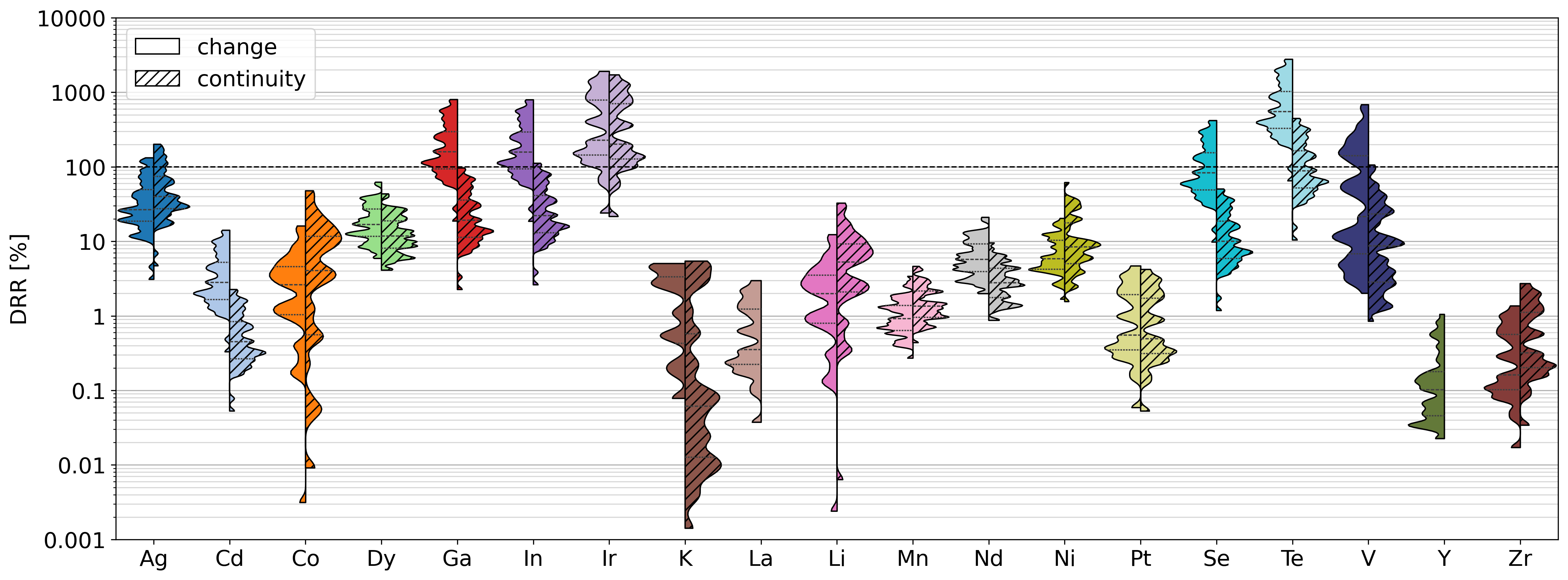}
        \caption{Change (left) vs continuity (right) sub-technology roadmap.}
        \label{fig:subtechno_roadmap}
    \end{subfigure}
    
    \caption{Distribution of demand-to-reserve ratio (DRR) across all studies with varying technological innovation.}
\end{figure}

Figure \ref{fig:material_intensity} compares DRRs with current to projected future material intensities (cf.~\hyperref[sec:METHODS]{Methods}).
Technological innovation in terms of decreasing material intensities can significantly reduce the demand of six materials (Ag, Ga, In, Ir, Se, Te) with DRRs above 100\%.
In particular Ag, In and Se demands decrease to below the reserves for most analysed studies.
This can be traced back to expected advances in material efficiencies for Ag use in crystalline silicon cells as well as In and Se use in thin-film PV technologies (Cdte, CIGS).
In contrast, despite expected improvements in material efficiency for Ir use in polymer electrolyte membrane electrolysis (PEMEL) \citep{wortmannCriticalIridiumDemands}, its DRR remains above 100\% for many studies. 
The DRRs of Te and V remain well above 100\% for a significant share of studies. 

Figure \ref{fig:subtechno_roadmap} compares the DRRs for two sub-technology roadmaps. The continuity roadmap largely assumes a conservative progress of current technological patterns, whereas the change roadmap assumes stronger market penetration of innovative sub-technologies \citep{schlichenmaierMayMaterialBottlenecks2022}. 
Here the trend is contrary: the continuity sub-technology roadmap has lower demand of many of the particularly critical materials than the change roadmap.
The increase in Ag and decrease in Ga, In, Se, Te demand are driven by a shift from thin-film PV technologies (copper indium gallium selenide and cadmium telluride solar cells cells) in the change roadmap towards the more mature cells (crystalline silicon) in the continuity roadmap. 
Because PEMEL are the only Ir demand driver and take a major share in both roadmaps its DRR is not significantly affected.
Y and La are only demanded in the change roadmap.
Y is used in high-temperature superconducting direct-drive wind turbines and high-temperature electrolysis (HTEL), whereas La is required only for HTEL. 
Taken together, technological innovation can significantly affect material demands, even to an extent that DRRs change by a factor of 2 to 7.
This effect may, however, go in any direction, leading to significantly increasing or decreasing demand for individual materials.

\section*{DISCUSSION AND CONCLUSIONS}

Our work questions neither the necessity nor the benefits of the energy transition towards renewables.
It stresses material demand as a challenge that needs to be tackled on the way.
The recent review by \citet{schulzeOvercomingChallengesAssessing2024} highlights a methodological gap in considering material demand in energy scenarios.
We went one step further and showed how this methodological gap regularly leads to model outcomes with material demands exceeding current reserves allocated to Europe.
Similar analyses of individual energy scenarios, e.g.\ by \citet{schlichenmaierMayMaterialBottlenecks2022}, similarly conclude that material bottlenecks are expected.

In particular, we found that the demand for Ag, Ga, In, Ir, Se, Te and V exceeds reserves in many highly decarbonised European energy system models.
This holds true even without competing non-energy material demand and especially when multiple energy sub-sectors are modelled.
The main technological drivers of these material demands are thin-film PV technologies such as copper indium gallium selenide (CIGS) and cadmium telluride (CdTe) solar cells for Ga, In, Se, Te, crystalline silicon PV for Ag and PEMEL for Ir demand.

We analysed the effect of technological innovation, namely a change sub-technology roadmap and decreasing material intensities, on material demands.
Both have a significant effect on material demands, but in opposite directions.
While the reduced material intensity decreases demands for several materials, the increased deployment of innovative sub-technologies, particularly thin-film PV (CIGS and CdTe), increases them.
Therefore, the question arises, what other measures may help addressing material shortages and ensure the feasibility of the energy transition.

\emph{First}, material demand in the energy sector may be reduced by lowering energy demand itself through efficiency or sufficiency measures.
Efficiency measures, however, are often accompanied by new equipment, e.g.\ electrical vehicles \citep{CriticalMaterialsBatteries}, transmission infrastructure \citep{MatReqElecGrids} or smart devices \citep{davidSmartNotSmart2019}, and therefore by new material demand.
Therefore, energy efficiency measures do not necessarily have a positive effect on material shortages, but may lead to new or intensified shortages, depending on the specific substitution.
Sufficiency measures, in contrast, that reduce energy service demands, can reduce both material and energy demands reliably, e.g.\ through lowering distance travelled or living area per person.
In light of difficulties to make the required changes in lifestyle or attitude \citep{michaelisSufficiencyStrategyEnough2024}, our analysis highlights the importance of energy sufficiency beyond climate neutrality and energy security \citep{wieseKeyRoleSufficiency2024}.

\emph{Second}, developing new (sub)-technologies may reduce material demand.
Our analysis shows for which materials and (sub)-technologies such a reduction or avoidance through innovation can be particularly effective.
New technologies may, however, not be cost competitive.
For instance, electromagnets may reduce the Nd demand of wind turbines compared to permanent magnets, but increase maintenance cost, especially for offshore farms \citep{RoleCriticalMinerals}.
Similarly, power-to-heat-to-power storages, so-called Carnot batteries, can have significantly lower material demand than lithium ion batteries through relying on more abundant materials such as concrete, water or molten salts.
But they have higher investment cost and lower efficiency \citep{nitschFutureRoleCarnot2024a}.
To lower material demand through such technological innovation, developers and investors would need to assess technologies not purely cost- or efficiency-based, but also regarding their material demand, ideally supported by policy.

\emph{Third}, material recycling or life time extension can address material shortages \citep{RecyclingCriticalMinerals}, at least if reserves suffice to build an initial technology fleet and capture as well as recycling rates are high.
Both measures, however, have a relevant time component, i.e.\ the materials of an initial fleet can be used longer or repeatedly.
The model results we analysed represent, in contrast, snapshots in time, not transition pathways.
Therefore, recycling or life time extension can only address the material shortages we identified, if sufficient amounts of the material are already in built stock.
This is questionable, for instance, for Ir, which has little competing non-energy demand and is mainly required for electrolysers that have little currently installed capacities.
Therefore, our analysis stresses that recycling and life time extensions cannot necessarily resolve shortages for all materials.

\emph{Fourth}, increasing reserves may alleviate material shortages. 
For instance, market prices and technological progress may drive the expansion of existing mining sites and the exploration of new occurrences to increase resources or increase the economically recoverable share of a resource. 
Such increases of reserves are, however, limited for Ga, In, Ir, Se and Te, as these raw materials are mainly recovered as by-products during the processing of other materials, while standalone recovery from primary deposits is often not economically viable \citep{MineralCommoditySummaries2026}.
Moreover, long lead times of new mining projects can delay reserve and production expansions \citep{buarqueandradeExplorationProductionUnderstanding2024}, so that an increase of reserves alone is not a reliable strategy to alleviate material shortages.

For energy system modellers, our results therefore indicate that neglecting material demand frequently leads to infeasible technology or infrastructure expansion plans, in particular for highly decarbonised multi-sectoral studies. Our study therefore calls for coupling detailed energy system models with material demand assessments and emphasises the need for methodological innovation highlighted by \citet{schulzeOvercomingChallengesAssessing2024}.

\newpage

\section*{METHODS}
\phantomsection
\label{sec:METHODS}

In this study we combine a systematic literature review of highly decarbonised energy system modelling (ESM) studies with a quantitative ex-post material assessment for 5 key technologies and 19 materials to analyse potential reserve constraints on modelled energy futures. First, we extract expansion capacities of  photovoltaic (PV), wind power, electrolysers, batteries, and  concentrated solar power (CSP) from 59 ESM studies and translate these capacities into material demands using technology-specific material intensities. Then we calculate the material demand-to-reserve ratio (DRR), which serves to identify the materials and technologies that may drive reserve constraints.

\subsection*{Systematic literature review of highly decarbonised European energy system optimisation studies}

\begin{figure}
    \centering
    \includegraphics[width=0.7\linewidth]{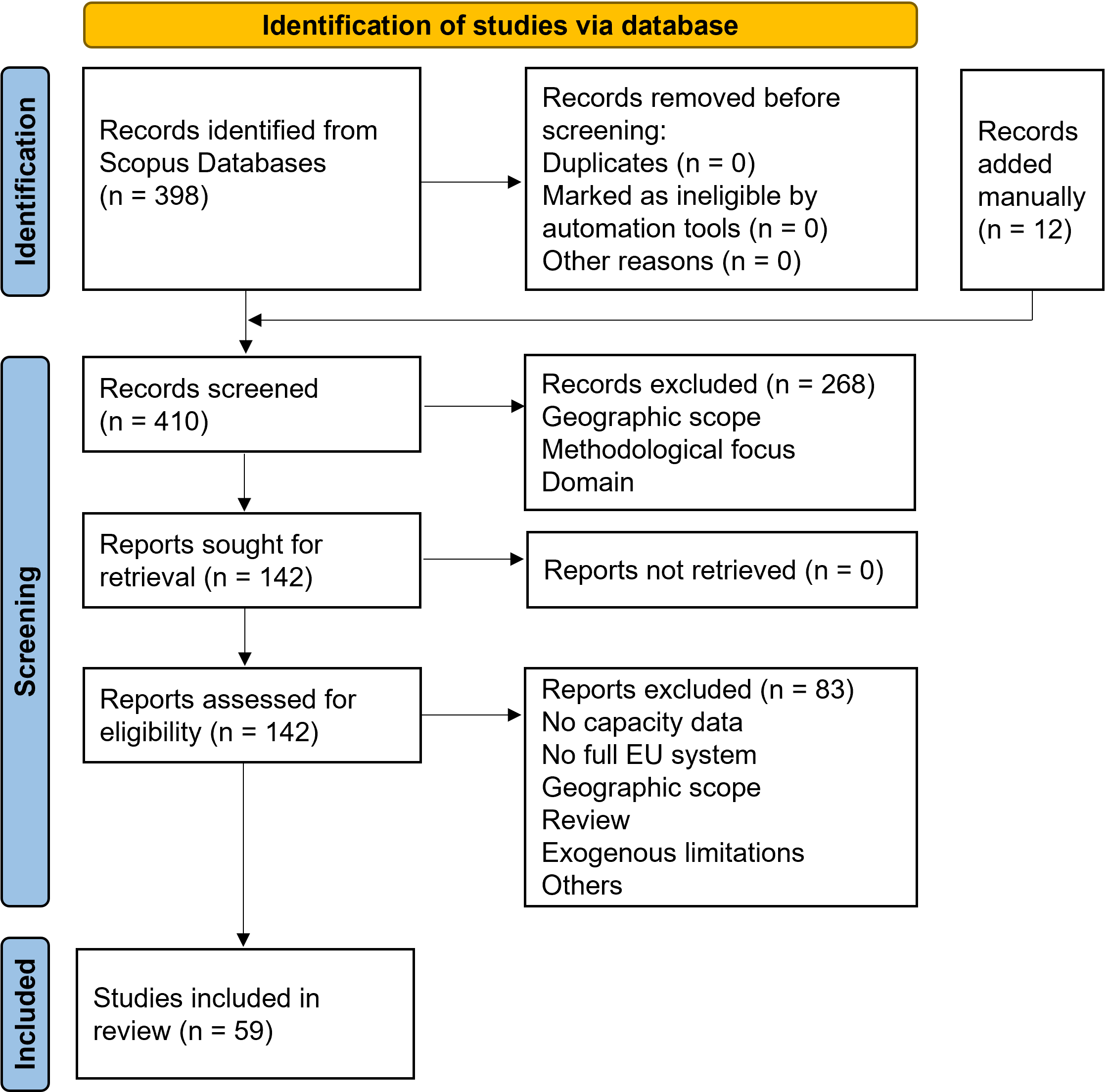}
    \caption{Literature review process based on the PRISMA guidelines.}
    \label{fig:PRISMA}
\end{figure}

Figure \ref{fig:PRISMA} illustrates the literature review process based on the PRISMA guidelines \citep{pagePRISMA2020Statement2021a}.
We first identify relevant papers, then snowballing and analysing their key terms to construct a search query aiming for highly decarbonised European energy system optimisation studies with endogenous capacity expansion.
The final search query is shown in Table \ref{tab:query}. 
Applying this query to the Scopus database \citep{Scopus} yielded 398 studies. 
We add 12 further studies not captured by the query to the initial population and conduct no grey literature search.

\input{Tables/table_query}

Manual screening reduces the initial population to 142 studies. Excluded studies typically feature mismatching geographic scopes (e.g.\ regional, national, global), have a strong methodological focus (e.g.\ technology learning), or focus specific topics (e.g.\ frequency control) not relevant for the scope of this study.
We retrieve all remaining studies in full text and assess eligibility by two independent reviewers based on predefined inclusion and exclusion criteria.
We consider studies eligible if they cover a European power system in a highly decarbonised future with endogenous capacity expansion.
We consider studies ineligible for the reasons summarised in Figure \ref{fig:PRISMA}. 
For a significant share of studies, this is due to the lack of transparent capacity data.
Applying all criteria results in a final population of 59 eligible studies. 

Many studies explore multiple energy system scenarios. To limit the scope of this review, we select one scenario per study. Where data transparency allowed, we choose the base case or reference scenario, in particular for studies using multi-objective optimisations, modelling to generate alternatives or stochastic optimisation.
Otherwise, we prefer higher levels of decarbonisation. 
Where studies model transition pathways we only use the year of highest decarbonisation as a snapshot. 

For the selected scenarios, we extract capacity data, preferably from numbers stated in tables or, where necessary and possible, from figures.
With the exception of batteries, we extract only capacities in terms of power rating, not energy supply.
As the technological details across ESMs are heterogeneous and the availability of material intensity data is limited, we focus on key technologies in decarbonised energy systems, namely PV, onshore and offshore wind, electrolysers, batteries and concentrated solar power.
In line with our lower boundary approach, neglecting other technologies makes our material demand estimates conservative.

We then harmonise the capacity data.
If studies distinguish central and distributed PV capacities, we sum these due to similar material demands\cite{RoleCriticalMinerals}.
If wind capacity is not classified as onshore or offshore, we disaggregate it using the mean onshore-to-offshore capacity ratio derived from all studies that make this distinction. 
Furthermore, only a minority of studies specifies sub-technologies.
We therefore neglect the few study-specific sub-technology assumptions and apply consistent sub-technology roadmaps across all studies (cf.~subsection \hyperref[sec:ex_post_material_assessment]{Ex-post material assessment} and Table~\ref{tab:sub-tech}).
Since battery material demand is proportional to energy capacity rather than power, we use a typical battery energy-to-power ratio of four hours where conversion is necessary \citep{bertschHowCanEnergysystem2025}. The final results of the literature review are presented in Figure \ref{fig:capacity} and Table \ref{tab:capacity}.

\input{Tables/table_capacities.tex}

\subsection*{Ex-post material assessment} 
\label{sec:ex_post_material_assessment}
\phantomsection
The ex-post material assessment broadly follows the approach from \citet{schlichenmaierMayMaterialBottlenecks2022}.
For each study, we calculate material Demand-to-Reserve Ratios for the 19 raw materials listed in Table \ref{tab:materials} that are required for the 5 key technologies and for which data are available.  We calculate the DRRs by combining the capacities with assumptions on future sub-technology market shares, corresponding materials intensities and current reserves as described below.

For each study $s$ and material $m$, we calculate the Demand-to-Reserve Ratio
\begin{equation}
    DRR_{s,m} = \frac{D_{s,m}^{\mathrm{tot}}}{R^{\mathrm{EU}}_m}
\end{equation}
as the ratio of total material demand $D_{s,m}^\mathrm{tot}$ and the allocated reserves $R^{\mathrm{EU}}_m$.
While $DRR_{s,m} > 100{\%}$ indicates a material shortage, $DRR_{s,m} < 100{\%}$ does not guarantee that a material is sufficiently available, given our lower boundary approach. Technologies not considered may induce additional material demands.
The material intensity of a technology depends on the underlying sub-technologies, which many studies do not report.
Therefore, we derive capacities $C_{s,t,u}$ per sub-technology $u$ of a technology $t$ from the extracted capacities $C_{s,t}$ for each study by applying projected market penetration shares $f_{t,u}$ for two future sub-technology roadmaps via
\begin{equation}
    C_{s,t,u} = f_{t,u} \cdot C_{\mathrm{s,t}}.
\end{equation}
The projected sub-technology market shares are stated in Table \ref{tab:sub-tech}.
Then we calculate the total demand $D_{s,m}^{\mathrm{tot}}$ for each material and study using material intensities specific to each sub-technology $I_{m,t,u}$ as
\begin{equation}
    D_{s,m}^{\mathrm{tot}} = \sum_{t} \sum_{u} C_{s,t,u} \cdot I_{m,t,u}.
\end{equation}
The material intensities are stated in Table \ref{tab:materialIntensity}.
Finally, we relate the total material demand to current reserves $R^{\mathrm{EU}}_m$. We assume that only a share $k$ of global reserves $R_{m}^{\mathrm{global}}$ is allocated to Europe proportional to either (a) its share of global population,  5.6\% \citep{kcWittgensteinCenterWIC2024}, or (b) its share of global GPD, 16.7\%, \citep{crespocuaresmaIncomeProjectionsClimate2017}, i.e.
\begin{equation}
    {R^{\mathrm{EU}}_m} = k \cdot R_{m}^{\mathrm{global}}.
\end{equation}

\subsection*{Sensitivity cases}

We conduct four sensitivity analyses.
The reference assumptions are indicated in bold font.
\begin{itemize}
    \item Reserve allocation (\textbf{population-based} vs GDP-based): As described above, global material reserves are allocated to Europe either based on a population-based fair-share approach, corresponding to 5.6 \% of global reserves, or according to a GDP-based allocation, corresponding to 16.7 \% of global reserves. Reserve data are based on \citet{MineralCommoditySummaries2026} where available. Remaining reserve data are based on \citet{schlichenmaierMayMaterialBottlenecks2022} (cf.\ Table \ref{tab:materials}).
    \item Material intensities (\textbf{2025} vs 2050): Material intensities of several sub-technologies are assumed to decrease over time due to technological progress. Current intensities are compared to a 2050 projection.  Both intensities data are based on \citet{schlichenmaierMayMaterialBottlenecks2022} (cf.\ Table \ref{tab:materialIntensity}) and interpolated where necessary.
    \item Sub-technology roadmaps (\textbf{``change''} vs ``continuation'' ): The continuation roadmap presumes the ongoing deployment of currently prevalent sub-technologies, whereas the change roadmap presumes a changing future market penetration of new sub-technologies (cf.\ Table \ref{tab:sub-tech}). From both roadmaps from \citet{schlichenmaierMayMaterialBottlenecks2022}, the sub-technology shares for the year 2050 are used.
    \item Competing material demand (\textbf{neglected} vs considered): A competing material demand from non-energy sectors $D^{\mathrm{NE}}_m$ may be considered (cf.\ Table \ref{tab:materials}) in addition to the demand from the extracted energy technology capacities via
    \begin{equation}
        DRR_{s,m}^* = \frac{D_{s, m}^{\mathrm{tot}} + D_{m}^{\mathrm{NE}}}{R_{m}^{\mathrm{EU}}}.
    \end{equation}
    $D_{m}^{\mathrm{NE}}$ is derived from \citet{schlichenmaierMayMaterialBottlenecks2022} by accumulating annual global non-energy material requirements over the period 2025 to 2050 and allocating the corresponding share to Europe using the population-based allocation approach described above.
\end{itemize}

\subsection*{Limitations}

Our study is subject to limitations related to the review process, the classification of studies and the assumptions made to estimate materials shortages.
In terms of the review process, the assessed studies often lack transparency concerning data and technology specifications. Some studies only report selected or aggregated results or do not state sub-technological specifications such as the modelled battery type. In addition, some data are available only in graphical form, making extraction prone to error. 

Since some studies apply decarbonisation budgets, whereas others refer to temperature targets, such as the \SI{1.5}{\celsius} warming limit, we only did a rough classification of studies in two decarbonisation categories (fully decarbonised vs others). Differences in the deployment of carbon capture and storage further constrain the comparability of the decarbonisation levels. Likewise, the twofold categorisation of sector coverage (electricity-only vs multi-sector) involves some ambiguity, since the degree of sector coupling varies substantially across the reviewed studies and the term itself is not consistently defined. Consequently, the multi-sector category covers a broad range of sector-coupling levels. 

Also the ex-post assessment of material demands is subject to limitations. Our study only identifies those material demands reliably that exceed allocated reserves. First, due to limited transparency of capacity data and heterogeneous technology coverage across models, we focus on five key technologies for decarbonised energy systems. Second, we consider material demands only for snapshots in time, rather than across full transition pathways. By excluding material demands from other energy technologies and from capacity replacement or decommissioning over time, our estimates are conservative. We therefore cannot claim that other materials are abundant.

Factors beyond reserves can constrain material availability. Even where reserves are sufficient, limited production can delay material supply, for instance as a results of high lead times for building up new mining projects. The availability of materials may also be constrained by geopolitical factors, in particular by the concentration of resources in specific countries, some of which are politically unstable, and the distribution of such resources within strategic trade partnerships. Additionally, many materials are traded as intermediate or finished products and hence limited by their upstream manufacturing that may take place elsewhere. In this study, we assumed a simplified allocation of global raw material reserves to Europe proportional to its population and GDP, respectively. While the former follows the idea of a globally fair distribution per head, the latter may be a better proxy to actual market access. Therefore, even materials with sufficient reserves are not necessarily available at the required time.

Finally, the quality of the data used to calculate the DRRs is limited. This applies in particular to the development of future material intensities and the assumed market penetration of the various sub-technologies. Technological progress may reduce material requirements, while disruptive advances may replace critical materials altogether. Conversely, the emergence of new sub-technologies may alter market shares and lead to different material demand patterns. For instance, recent advances in battery technologies have reduced cobalt intensities, illustrating how future developments may substantially affect material demand estimates \citep{soltzerMaterialBottlenecksBatteries2026}.

\newpage

\section*{ACKNOWLEDGEMENTS}
This work was partly funded by the European Union (ERC, MATERIALIZE, 101076649). Views and opinions expressed are however those of the authors only and do not necessarily reflect those of European Union or the European Research Council Executive Agency. Neither the European Union nor the granting authority can be held responsible for them.
This work was supported by the Helmholtz Association under the program “Energy System Design.” 
This work was partly funded by the Deutsche Forschungsgemeinschaft (DFG, German Research Foundation) -- 526062606 within the priority programme ``Carnot Batteries: Inverse Design from Markets to Molecules'' (SPP 2403).

\section*{AUTHOR CONTRIBUTIONS}
J.M.: Conceptualisation, Methodology, Investigation, Writing-–original draft, Writing-–review \& editing, Visualisation, Formal analysis, Data curation
J.F.: Conceptualisation, Methodology, Writing-–original draft, Writing-–review \& editing,
K.E.: Investigation, Writing-–review \& editing,
H.H.: Writing-–review \& editing, funding acquisition, supervision, conceptualisation

\section*{DECLARATION OF INTERESTS}
The authors declare no competing interests.

\section*{DECLARATION OF GENERATIVE AI AND AI-ASSISTED TECHNOLOGIES}

During the preparation of this work, the authors used ChatGPT (GPT-5) and DeepL Write in
order to improve clarity and language. After using this tool/service, the authors reviewed and
edited the content as needed and take full responsibility for the content of the published article.

\newpage

\bibliography{references_V3.bib}

\newpage

\section*{APPENDIX}

\subsection*{Further result figures}

\begin{figure}[htpb]
    \centering
    \includegraphics[width=1\linewidth]{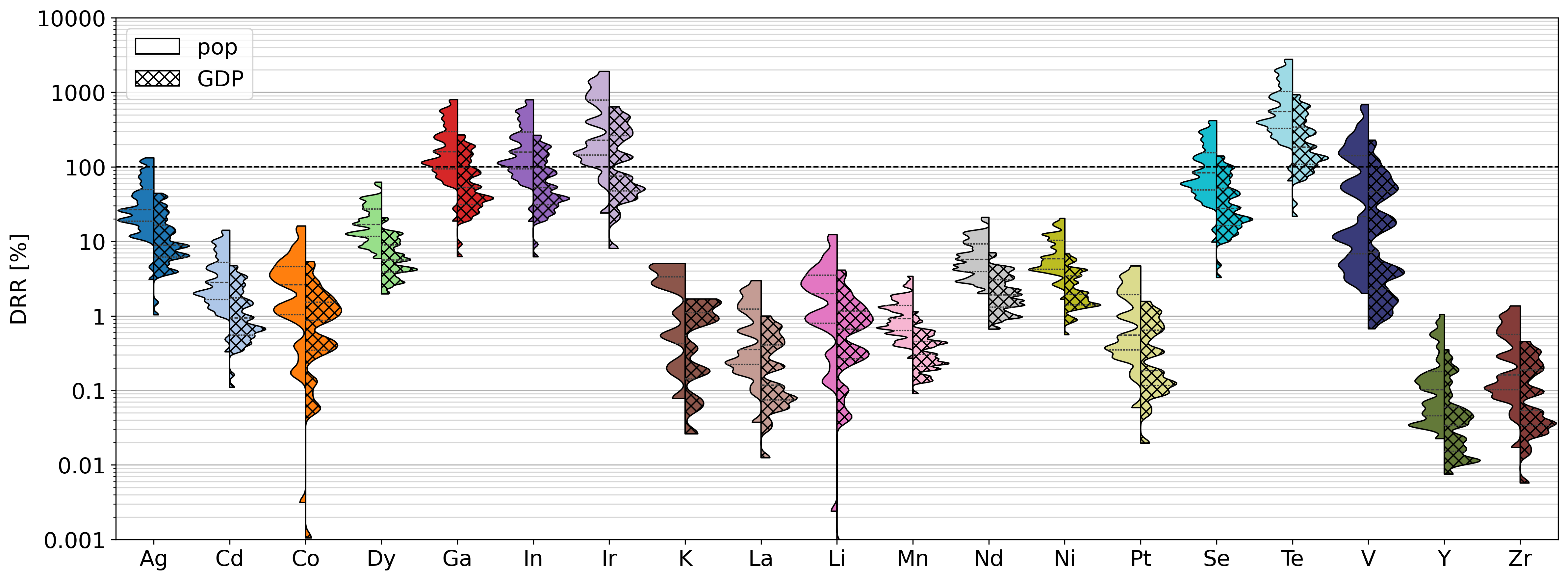}
    \caption{Distribution of demand-to-reserve ratio (DRR) across all studies with population-based (left) vs GDP-based (right) reserve allocation.}
    \label{fig:gdp_allocation}
\end{figure}

\begin{figure}[htpb]
    \centering
    \includegraphics[width=1\linewidth]{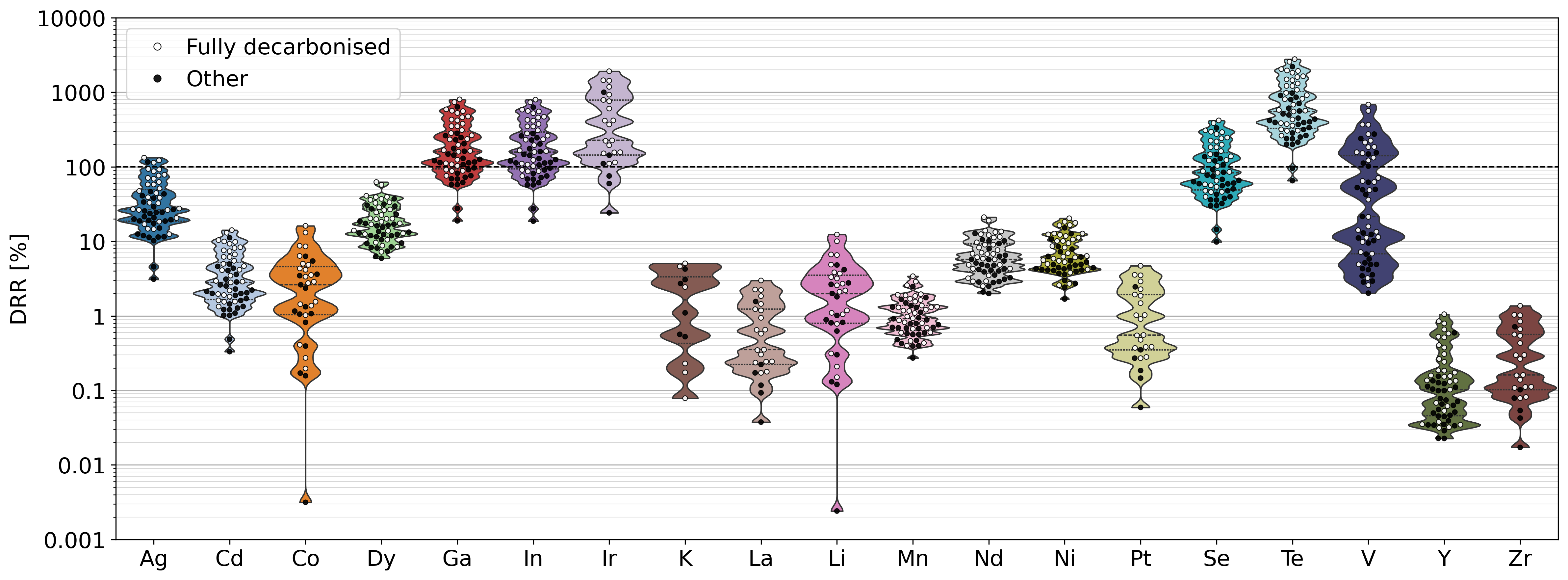}
    \caption{Distribution of demand-to-reserve ratio (DRR) across all studies in the reference case, where white markers indicate the fully decarbonised scenarios.}
    \label{fig:decarb}
\end{figure}

Figure \ref{fig:gdp_allocation} compares population-based vs GDP-based reserve allocation.
Figure \ref{fig:decarb} indicates the distribution of studies according to decarbonisation level.

\subsection*{Further input data tables}

Table \ref{tab:materials} summarises the current global reserves and competing non-energy demands for all 19 materials analysed.
Table \ref{tab:sub-tech} indicates the sub-technology market shares and Table \ref{tab:materialIntensity} the material intensities for all materials and sub-technologies.

\input{Tables/table_materials}

\input{Tables/table_market_shares}

\input{Tables/table_material_intensity}

\end{document}

%% file: Tables/table_query.tex
\begin{table}[htpb]
    \centering
    \caption{Search query applied in Scopus. All sub-queries are combined using AND.}
    \label{tab:query}
    \footnotesize
    \begin{tabular}{p{2.8cm} p{11.5cm}}
    \toprule
        Field Code & Sub-query \\
    \midrule
        TITLE-ABS-KEY&((europ* AND (("energy system" OR "electricity system" OR "power system") W/2 ("model*" OR "optim*" OR "design*"))) OR ((europ* W/2 ("energy system" OR "electricity system" OR "power system")) AND ("model*" OR "optim*" OR "design*")))\\
        TITLE-ABS-KEY & ("system cost" OR "total cost" OR "overall costs" OR "least cost" OR "cost effective" OR "optimal cost" OR "cost optim*“ OR "cost minim*" OR “minim* cost” OR "capacity expansion" OR "capacity investment" OR “system investment” OR "investment strategy" OR "strategic investment" OR "expansion model*" OR "strategic plan*" OR "capacity plan*" OR "long term plan*" OR "investment plan*“ OR "infrastructure plan*" 	OR	"layout" OR "system configuration" OR "infrastructure design" OR "infrastructure requirements" OR "optim* design“ OR “system design” OR “optimal system” OR "decarboni?ation scenario" OR "decarboni?ation strategy“ OR “long term scenario” OR “transition pathway”)\\
        PUBYEAR& $>$ 2014\\
        PUBYEAR& $<$ 2027\\
        EXCLUDE& (DOCTYPE,"cp" )\\
    \bottomrule
    \end{tabular}
\end{table}

%% file: Tables/table_capacities.tex
\begin{table}
\caption{Capacities in GW, Battery capacities in GWh, ESM name, sectoral scope and decarbonisation level extracted from the 59 studies resulting from the systematic literature review. The same data is visualised in Figure \ref{fig:capacity}.}
\label{tab:capacity}
\scriptsize
\begin{tabular}{
r
*{6}{S[round-mode=places, round-precision=0, table-format=4.0]}
llll
}
\toprule
ID
& \multicolumn{1}{c}{PV}
& \multicolumn{1}{c}{CSP}
& \multicolumn{1}{c}{ONWIND}
& \multicolumn{1}{c}{OFFWIND}
& \multicolumn{1}{c}{BATTERY}
& \multicolumn{1}{c}{ELECTROLYSER}
& ESM name & sector & decarbonisation & reference \\
\midrule
1 & 4695.00 & 0.00 & 1274.80 & 142.70 & 1843.00 & 1653.70 & PyPSA & multi & fully & \citep{pedersenLongtermImplicationsReduced2022} \\
2 & 5100.00 & 0.00 & 1141.00 & 159.00 & 1348.00 & 1349.00 & PyPSA & multi & fully & \citep{rahdanStrategicDeploymentSolar2025} \\
3 & 3351.00 & 0.00 & 1785.00 & 175.00 & 5000.00 & 2186.00 & PyPSA & multi & fully & \citep{zeyenEndogenousLearningGreen2023} \\
4 & 4040.00 & 121.00 & 1583.00 & 69.00 & 2076.00 & 1142.00 & PyPSA & multi & other & \citep{rahdanDistributedPhotovoltaicsProvides2024} \\
5 & 3747.00 & 182.00 & 1622.85 & 530.15 & 1912.00 & 866.00 & PyPSA & multi & fully & \citep{gotskeDesigningSectorcoupledEuropean2024} \\
6 & 3592.69 & 0.00 & 1696.27 & 212.03 & 1093.48 & 1060.05 & PyPSA & multi & fully & \citep{neumannPotentialRoleHydrogen2023} \\
7 & 2661.00 & 0.00 & 2972.00 & 0.00 & 156.00 & 478.00 & Calliope & multi & fully & \citep{pickeringDiversityOptionsEliminate2022} \\
8 & 2225.00 & 0.00 & 1742.00 & 464.00 & 0.00 & 1629.00 & Enertile & multi & fully & \citep{luxSupplyCurveElectricitybased2020} \\
9 & 2230.00 & 0.00 & 2909.00 & 411.00 & 0.00 & 0.00 & Calliope & multi & fully & \citep{wuStrategicUsesAncillary2023} \\
10 & 2730.00 & 0.00 & 1371.85 & 448.15 & 1280.00 & 900.00 & GENeSYS & elec & fully & \citep{bussarLargescaleIntegrationRenewable2016} \\
11 & 3543.60 & 0.00 & 872.80 & 194.50 & 0.00 & 254.60 & COMPETES & multi & fully & \citep{morales-espanaImpactLargescaleHydrogen2024} \\
12 & 2934.30 & 0.00 & 1304.40 & 223.70 & 0.00 & 172.80 & urbs & multi & fully & \citep{gawlickImpactCouplingElectricity2023} \\
13 & 2405.00 & 0.00 & 1078.63 & 352.37 & 1584.00 & 475.00 & PyPSA & multi & fully & \citep{akhmetovFlatteningPeakDemand2025} \\
14 & 3008.00 & 0.00 & 535.00 & 234.00 & 0.00 & 179.00 & PyPSA & multi & fully & \citep{victoriaEarlyDecarbonisationEuropean2020} \\
15 & 1479.76 & 0.00 & 1345.31 & 439.49 & 0.00 & 0.00 & AnyMOD & multi & fully & \citep{aliasghariPotentialElectrifiedTransport2025} \\
16 & 1499.48 & 0.00 & 1035.62 & 166.56 & 3261.22 & 422.27 & FINE & multi & fully & \citep{dunkelHydrogenAutarkyEvaluating} \\
17 & 1787.00 & 0.00 & 877.00 & 392.00 & 312.00 & 0.00 & ISAaR  & multi & other & \citep{kernModelingEvaluatingBidirectionally2022} \\
18 & 1797.30 & 6.88 & 917.33 & 151.72 & 6153.55 & 178.00 & Backbone & multi & fully & \citep{johanndeiterPriceFormationIntersectoral2024} \\
19 & 939.00 & 0.00 & 1527.12 & 498.88 & 0.00 & 0.00 & E2M2s & multi & other & \citep{blumbergImpactElectricVehicles2022} \\
20 & 1677.69 & 9.06 & 944.86 & 127.55 & 2429.50 & 125.99 & Backbone & multi & fully & \citep{johanndeiterDoesDifferenceMake2025} \\
21 & 1038.10 & 0.00 & 755.80 & 369.20 & 0.00 & 690.00 & MIRET-EU & multi & fully & \citep{seckHydrogenDecarbonizationEnergy2022} \\
22 & 1665.00 & 0.00 & 819.70 & 176.90 & 505.20 & 0.00 & ELTRAMOD & multi & other & \citep{misconelAssessingValueDemand2021} \\
23 & 556.40 & 3.10 & 1737.90 & 364.10 & 0.00 & 0.00 & no name & elec & fully & \citep{dominguezAnalysingDecarbonizingStrategies2021} \\
24 & 1975.00 & 0.00 & 546.00 & 0.00 & 3320.00 & 0.00 & LUT & elec & fully & \citep{childFlexibleElectricityGeneration2019} \\
25 & 909.00 & 0.00 & 1119.34 & 365.66 & 0.00 & 0.00 & elesplan-m & elec & other & \citep{plessmannHowMeetEU2017} \\
26 & 722.00 & 0.00 & 1218.08 & 397.92 & 2400.00 & 0.00 & MIM & multi & other & \citep{yueksel-erguenModelingTransitionMultimodal2025} \\
27 & 719.00 & 0.00 & 1567.00 & 21.00 & 0.00 & 0.00 & PyPSA & multi & other & \citep{zhuImpactCO2Prices2019} \\
28 & 1296.30 & 0.00 & 738.30 & 121.30 & 402.40 & 0.00 & SCOPE & multi & other & \citep{hartelAggregationMethodsModelling2017} \\
29 & 1572.00 & 0.00 & 341.00 & 53.00 & 1380.00 & 0.00 & REMix & multi & other & \citep{junneConsideringLifeCycle2021} \\
30 & 1462.00 & 0.00 & 485.00 & 0.00 & 1320.00 & 0.00 & OSeMOSYS & elec & other & \citep{heggartyAssessingRelativeImpacts2024} \\
31 & 1026.48 & 0.00 & 69.25 & 708.76 & 0.00 & 126.00 & DIMENSION & multi & other & \citep{helgesonRoleElectricityDecarbonizing2020} \\
32 & 800.00 & 0.00 & 197.00 & 872.00 & 0.00 & 0.00 & EMPIRE & elec & other & \citep{backeEMPIREOpensourceModel2022} \\
33 & 654.00 & 0.00 & 842.00 & 78.00 & 590.00 & 258.00 & FINE & multi & fully & \citep{caglayanRobustDesignFuture2021} \\
34 & 707.00 & 0.00 & 774.00 & 127.00 & 389.60 & 222.00 & Backbone & elec & fully & \citep{mutkeInfluenceBioenergyTransmission2023} \\
35 & 1012.10 & 0.00 & 491.40 & 280.80 & 522.40 & 0.00 & IESA-EUPS & elec & fully & \citep{beresImpactNationalPolicies2024} \\
36 & 1250.00 & 0.00 & 500.00 & 0.00 & 0.00 & 0.00 & PyPSA & multi & fully & \citep{vangreevenbroekTradingRegionalOverall2025} \\
37 & 686.00 & 96.20 & 625.40 & 47.40 & 546.80 & 130.60 & REMix & multi & fully & \citep{sasanpourStrategicPolicyTargets2021} \\
38 & 1120.00 & 0.00 & 340.21 & 111.14 & 899.20 & 0.00 & PyPSA & elec & other & \citep{pedersenModelingAllAlternative2021} \\
39 & 766.70 & 0.00 & 800.00 & 0.00 & 0.00 & 0.00 & EMPIRE & elec & other & \citep{maranon-ledesmaAnalyzingDemandResponse2019} \\
40 & 1060.00 & 0.00 & 346.73 & 113.27 & 1650.00 & 0.00 & PyPSA & elec & fully & \citep{cheAssessSpacebasedSolar2025} \\
41 & 791.00 & 0.00 & 488.00 & 198.00 & 104.00 & 0.00 & PLEXOS & multi & fully & \citep{beresAssessingFeasibilityCO22024} \\
42 & 825.00 & 0.00 & 570.00 & 59.00 & 441.60 & 0.00 & PyPSA & elec & other & \citep{victoriaRolePhotovoltaicsSustainable2020} \\
43 & 480.36 & 0.00 & 911.33 & 49.51 & 1050.84 & 0.00 & PyPSA & elec & fully & \citep{neumannBroadRangesInvestment2023} \\
44 & 482.45 & 0.00 & 796.34 & 104.63 & 59.98 & 0.00 & EMPIRE & elec & other & \citep{hjelmelandRoleNuclearEnergy2025} \\
45 & 438.78 & 0.00 & 323.98 & 448.98 & 0.00 & 163.27 & PyPSA & elec & other & \citep{neumannNearoptimalFeasibleSpace2021} \\
46 & 436.00 & 0.00 & 629.00 & 280.00 & 0.00 & 0.00 & EMPIRE & elec & other & \citep{backeStableStochasticCapacity2021} \\
47 & 679.00 & 20.76 & 537.80 & 74.75 & 65.00 & 0.00 & TIMES & multi & other & \citep{ringkjobShorttermSolarWind2020} \\
48 & 580.00 & 0.00 & 545.00 & 119.00 & 408.00 & 68.00 & PyPSA & elec & other & \citep{schlachtbergerCostOptimalScenarios2018} \\
49 & 392.00 & 0.00 & 328.00 & 464.00 & 150.00 & 86.00 & REMix & elec & other & \citep{cebullaElectricalEnergyStorage2017} \\
50 & 615.00 & 107.40 & 397.00 & 118.00 & 1.20 & 0.00 & REMix & multi & other & \citep{caoPreventPromoteGrid2021} \\
51 & 520.56 & 43.38 & 490.47 & 160.23 & 0.00 & 0.00 & WeSim & elec & other & \citep{tengAssessmentFutureWholeSystem2018} \\
52 & 581.75 & 200.00 & 271.50 & 128.18 & 0.00 & 0.00 & PLEXOS & multi & fully & \citep{zappa100RenewableEuropean2019} \\
53 & 736.00 & 0.00 & 281.00 & 33.00 & 0.00 & 0.00 & TIMES & elec & other & \citep{gotzensInfluenceContinuedReductions2018} \\
54 & 459.30 & 0.00 & 509.30 & 45.10 & 0.00 & 27.50 & DIMENSION & elec & other & \citep{peterHowDoesClimate2019} \\
55 & 560.65 & 0.00 & 445.07 & 12.10 & 74.90 & 0.00 & Calliope & elec & fully & \citep{trondleTradeOffsGeographicScale2020} \\
56 & 367.00 & 168.00 & 327.00 & 0.00 & 996.00 & 0.00 & PyPSA & elec & other & \citep{duboisComputingNecessaryConditions2022} \\
57 & 366.07 & 22.32 & 325.89 & 118.30 & 0.00 & 0.00 & LIMES-EU+ & elec & other & \citep{schmidQuantifyingLongtermEconomic2015} \\
58 & 119.42 & 0.00 & 669.42 & 0.00 & 0.00 & 0.00 & PyPSA & elec & other & \citep{schlottImpactClimateChange2018} \\
59 & 175.00 & 0.00 & 345.22 & 112.78 & 0.00 & 0.00 & WeSIM & multi & other & \citep{papadaskalopoulosQuantifyingPotentialEconomic2018} \\
\bottomrule
\end{tabular}
\end{table}

%% file: Tables/table_materials.tex
\begin{table}[htpb]
    \centering
    \caption{Material abbreviation, full material name, current global reserves and competing non-energy demands for all 19 materials analysed. The EU is allocated 5.6\% of the reserves for population-based allocation, and 16.7\% for GDP-based allocation. Reserve data are based on \citet{MineralCommoditySummaries2026} where available. Remaining reserve data and competing non-energy demand is based on \citet{schlichenmaierMayMaterialBottlenecks2022}.
    \label{tab:materials}}
    \footnotesize
    \begin{tabular}{l l r r r}
    \toprule
        Abbreviation & Material & Current global reserves (kt) & Competing non-energy demand (kt) & Coverage (\%) \\
    \midrule
        Ag& Silver & 610.0 & 1721.0 & 282.1 \\
        Cd& Cadmium & 7200.0 & 2170.0 & 30.1 \\
        Co& Cobalt & 12000.0 & 8665.1 & 72.2 \\
        Dy& Dysprosium & 544.0 & 51.3 & 9.4 \\
        Ga& Gallium & 5.2 & 25.9 & 497.6 \\
        In& Indium & 47.1 & 53.9 & 114.5 \\
        Ir& Iridium & 0.4 & 0.4 & 85.9 \\
        K& Potassium & 4897876.9 & 1621394.5 & 33.1 \\
        La& Lanthanum & 25330.0 & 2245.2 & 8.9 \\
        Li& Lithium & 37000.0 & 3073.3 & 8.3 \\
        Mn& Manganese & 801000.0 & 1437631.1 & 179.5 \\
        Nd& Neodymium & 16070.0 & 1389.7 & 8.6 \\
        Ni& Nickel & 140000.0 & 194503.5 & 138.9 \\
        Pt& Platinum & 44.0 & 17.2 & 39.2 \\
        Se& Selenium & 99.0 & 132.9 & 134.3 \\
        Te& Tellurium & 35.0 & 16.5 & 47.2 \\
        V& Vanadium & 21000.0 & 6103.6 & 29.1 \\
        Y& Yttrium & 3120.0 & 452.0 & 14.5 \\
        Zr& Zirconium & 51800.0 & 52496.0 & 101.3 \\
    \bottomrule
    \end{tabular}
\end{table}

%% file: Tables/table_market_shares.tex
\begin{table}[htpb]
    \centering
    \caption{Projected sub-technology market shares in 2050 derived from \citet{schlichenmaierMayMaterialBottlenecks2022}.}
    \scriptsize
    \begin{tabular}{l l l l r}
    \toprule
       Scenario & Technology & Sub-technology abbrev. & Sub-technology & $f_{t,u}$ (\%) \\
    \midrule
       continuity & PV & CIGS & Copper-Indium-Gallium-Diselenide & 1.5 \\
       continuity & PV & CdTe & Cadmium-Telluride & 4.0 \\
       continuity & PV & c-Si & Crystalline Silicon & 94.5 \\
       change (ref.) & PV & CIGS & Copper-Indium-Gallium-Diselenide & 12.5 \\
       change (ref.) & PV & CdTe & Cadmium-Telluride & 25.0 \\
       change (ref.) & PV & c-Si & Crystalline Silicon & 62.5 \\
       continuity & ONWIND &  HTS-DD & High temperature superconductor - direct drive & 0.0 \\
       continuity & ONWIND &  SG-PM-DD & Synchronous generator - permanent magnet - direct drive & 0.0 \\
       continuity & ONWIND &  SG-PM-MS & Synchronous generator - permanent magnet - middle spead gear & 5.0 \\
       continuity & ONWIND &  SG-PM-HS & Synchronous generator - permanent magnet - high speed gear & 28.0 \\
       continuity & ONWIND &  SG-E-DD & Synchronous generator - electrically excited - direct drive & 55.0 \\
       continuity & ONWIND &  AG & Asynchronous generator - electrically excited - high speed gear & 12.0 \\
       change (ref.) & ONWIND &  HTS-DD & High temperature superconductor - direct drive & 12.0 \\
       change (ref.) & ONWIND &  SG-PM-DD & Synchronous generator - permanent magnet - direct drive & 28.0 \\
       change (ref.) & ONWIND &  SG-PM-MS & Synchronous generator - permanent magnet - middle spead gear & 13.0 \\
       change (ref.) & ONWIND &  SG-PM-HS & Synchronous generator - permanent magnet - high speed gear & 42.0 \\
       change (ref.) & ONWIND &  SG-E-DD & Synchronous generator - electrically excited - direct drive & 3.0 \\
       change (ref.) & ONWIND &  AG & Asynchronous generator - electrically excited - high speed gear & 2.0 \\
       continuity & OFFWIND &  HTS-DD & High temperature superconductor - direct drive & 0.0 \\
       continuity & OFFWIND &  SG-PM-DD & Synchronous generator - permanent magnet - direct drive & 14.0 \\
       continuity & OFFWIND &  SG-PM-MS & Synchronous generator - permanent magnet - middle spead gear & 34.0 \\
       continuity & OFFWIND &  SG-PM-HS & Synchronous generator - permanent magnet - high speed gear & 2.0 \\
       continuity & OFFWIND &  AG & Asynchronous generator - electrically excited - high speed gear & 50.0 \\
       change (ref.) & OFFWIND &  HTS-DD & High temperature superconductor - direct drive & 17.0 \\
       change (ref.) & OFFWIND &  SG-PM-DD & Synchronous generator - permanent magnet - direct drive & 20.0 \\
       change (ref.) & OFFWIND &  SG-PM-MS & Synchronous generator - permanent magnet - middle spead gear & 61.0 \\
       change (ref.) & OFFWIND &  SG-PM-HS & Synchronous generator - permanent magnet - high speed gear & 0.0 \\
       change (ref.) & OFFWIND &  SG-E-DD & Synchronous generator - electrically excited - direct drive & 0.0 \\
       change (ref.) & OFFWIND &  AG & Asynchronous generator - electrically excited - high speed gear & 2.0 \\
       continuity & BATTERY & NMC-622 & Lithium-Nickel-Manganese-Cobalt-Oxide (Ni-Mn-Co = 6-2-2) & 6.0 \\
       continuity & BATTERY & NMC-811 & Lithium-Nickel-Manganese-Cobalt-Oxide (Ni-Mn-Co = 8-1-1) & 44.0 \\
       continuity & BATTERY & LFP & Lithium-Iron-Phosphate & 40.0 \\
       continuity & BATTERY & Redox-Flow & Vanadium-Redox-Flow & 5.0 \\
       continuity & BATTERY & NaS & Natrium-Sulfur & 5.0 \\
       change (ref.) & BATTERY & NMC-622 & Lithium-Nickel-Manganese-Cobalt-Oxide (Ni-Mn-Co = 6-2-2) & 2.0 \\
       change (ref.) & BATTERY & NMC-811 & Lithium-Nickel-Manganese-Cobalt-Oxide (Ni-Mn-Co = 8-1-1) & 15.0 \\
       change (ref.) & BATTERY & LFP & Lithium-Iron-Phosphate & 17.0 \\
       change (ref.) & BATTERY & Redox-Flow & Vanadium-Redox-Flow & 33.0 \\
       change (ref.) & BATTERY & NaS & Natrium-Sulfur & 33.0 \\
       continuity & ELECTROLYSER & AEL & Alkaline Water Electrolysis & 33.0 \\
       continuity & ELECTROLYSER & HTEL-Y & High-Temperature Electrolysis - Yttrium & 0.0 \\
       continuity & ELECTROLYSER & PEMEL & Polymer Electrolyte Membrane Electrolysis & 67.0 \\
       change (ref.) & ELECTROLYSER & AEL & Alkaline Water Electrolysis & 0.0 \\
       change (ref.) & ELECTROLYSER & HTEL-Y & High-Temperature Electrolysis - Yttrium & 25.0 \\
       change (ref.) & ELECTROLYSER & PEMEL & Polymer Electrolyte Membrane Electrolysis & 75.0 \\
       continuity & CSP & - & Parabolic Trough & 90.0 \\
       continuity & CSP & - & Fresnel Collector & 5.0 \\
       continuity & CSP & - & Solar Tower & 5.0 \\
       change (ref.) & CSP & - & Parabolic Trough & 42.5 \\
       change (ref.) & CSP & - & Fresnel Collector & 15.0 \\
       change (ref.) & CSP & - & Solar Tower & 42.5 \\
    \bottomrule
    \end{tabular}
    \label{tab:sub-tech}
\end{table}

%% file: Tables/table_material_intensity.tex
\begin{sidewaystable}[htpb]
    \centering
    \captionsetup{skip=2pt}
    \caption{Material intensities for all considered technologies in t/GW. Battery material intensities in t/GWh. Based on \citet{schlichenmaierMayMaterialBottlenecks2022}.}
    \scriptsize
    \setlength{\tabcolsep}{3.0pt}
    \providecommand{\miNA}{--}
    \newlength{\miMaterialColWidth}
    \settowidth{\miMaterialColWidth}{423.50}
    \begin{tabular}{
    @{}lll*{21}{p{\miMaterialColWidth}}@{}
    }
    \toprule
    Year & Tech & Sub-tech
    & \multicolumn{1}{l}{Ag}
    & \multicolumn{1}{l}{Cd}
    & \multicolumn{1}{l}{Co}
    & \multicolumn{1}{l}{Dy}
    & \multicolumn{1}{l}{Ga}
    & \multicolumn{1}{l}{Ge}
    & \multicolumn{1}{l}{In}
    & \multicolumn{1}{l}{Ir}
    & \multicolumn{1}{l}{K}
    & \multicolumn{1}{l}{La}
    & \multicolumn{1}{l}{Li}
    & \multicolumn{1}{l}{Mn}
    & \multicolumn{1}{l}{Nd}
    & \multicolumn{1}{l}{Ni}
    & \multicolumn{1}{l}{Pt}
    & \multicolumn{1}{l}{S}
    & \multicolumn{1}{l}{Se}
    & \multicolumn{1}{l}{Te}
    & \multicolumn{1}{l}{V}
    & \multicolumn{1}{l}{Y}
    & \multicolumn{1}{l}{Zr} \\
    \midrule
    2025 & PV & c-Si & 14.1 & \miNA & \miNA & \miNA & \miNA & \miNA & \miNA & \miNA & \miNA & \miNA & \miNA & \miNA & \miNA & 1.210 & \miNA & \miNA & \miNA & \miNA & \miNA & \miNA & \miNA \\
    2025 & PV & a-Si & \miNA & \miNA & \miNA & \miNA & \miNA & 35.4 & \miNA & \miNA & \miNA & \miNA & \miNA & \miNA & \miNA & 334.0 & \miNA & \miNA & \miNA & \miNA & \miNA & \miNA & \miNA \\
    2025 & PV & CIGS & \miNA & 0.88 & \miNA & \miNA & 3.64 & \miNA & 16.4 & \miNA & \miNA & \miNA & \miNA & \miNA & \miNA & \miNA & \miNA & \miNA & 36.2 & \miNA & \miNA & \miNA & \miNA \\
    2025 & PV & CdTe & \miNA & 43.9 & \miNA & \miNA & \miNA & \miNA & 8.17 & \miNA & \miNA & \miNA & \miNA & \miNA & \miNA & \miNA & \miNA & \miNA & \miNA & 42.5 & \miNA & \miNA & \miNA \\
    2025 & CSP & Parabolic Trough & 25.2 & \miNA & \miNA & \miNA & \miNA & \miNA & \miNA & \miNA & 74020 & \miNA & \miNA & 2000 & \miNA & 940.0 & \miNA & \miNA & \miNA & \miNA & 1.900 & \miNA & \miNA \\
    2025 & CSP & Fresnel Collector & 30.6 & \miNA & \miNA & \miNA & \miNA & \miNA & \miNA & \miNA & 85100 & \miNA & \miNA & 2000 & \miNA & 940.0 & \miNA & \miNA & \miNA & \miNA & 1.900 & \miNA & \miNA \\
    2025 & CSP & Solar Tower & 35.1 & \miNA & \miNA & \miNA & \miNA & \miNA & \miNA & \miNA & 58000 & \miNA & \miNA & 5700 & \miNA & 1800 & \miNA & \miNA & \miNA & \miNA & 1.700 & \miNA & \miNA \\
    2025 & EL & AEL & \miNA & \miNA & 6.70 & \miNA & \miNA & \miNA & \miNA & \miNA & 423.50 & \miNA & \miNA & \miNA & \miNA & 2922 & \miNA & \miNA & \miNA & \miNA & \miNA & \miNA & 108 \\
    2025 & EL & PEMEL & \miNA & \miNA & \miNA & \miNA & \miNA & \miNA & \miNA & 0.286 & \miNA & \miNA & \miNA & \miNA & \miNA & \miNA & 0.070 & \miNA & \miNA & \miNA & \miNA & \miNA & \miNA \\
    2025 & EL & HTEL\_SOEL-Y & \miNA & \miNA & \miNA & \miNA & \miNA & \miNA & \miNA & \miNA & \miNA & 76.8 & \miNA & \miNA & \miNA & 193.2 & \miNA & \miNA & \miNA & \miNA & \miNA & 2.89 & 71.9 \\
    2025 & BAT & NMC-111 & \miNA & \miNA & 407 & \miNA & \miNA & \miNA & \miNA & \miNA & \miNA & \miNA & 149 & 361.9 & \miNA & 414.0 & \miNA & \miNA & \miNA & \miNA & \miNA & \miNA & \miNA \\
    2025 & BAT & NMC-622 & \miNA & \miNA & 182 & \miNA & \miNA & \miNA & \miNA & \miNA & \miNA & \miNA & 132 & 172.8 & \miNA & 551.0 & \miNA & \miNA & \miNA & \miNA & \miNA & \miNA & \miNA \\
    2025 & BAT & NMC-811 & \miNA & \miNA & 92.3 & \miNA & \miNA & \miNA & \miNA & \miNA & \miNA & \miNA & 114 & 86.50 & \miNA & 739.2 & \miNA & \miNA & \miNA & \miNA & \miNA & \miNA & \miNA \\
    2025 & BAT & LFP & \miNA & \miNA & \miNA & \miNA & \miNA & \miNA & \miNA & \miNA & \miNA & \miNA & 127 & \miNA & \miNA & 64.50 & \miNA & \miNA & \miNA & \miNA & \miNA & \miNA & \miNA \\
    2025 & BAT & Redox-Flow & \miNA & \miNA & \miNA & \miNA & \miNA & \miNA & \miNA & \miNA & \miNA & \miNA & \miNA & \miNA & \miNA & \miNA & \miNA & \miNA & \miNA & \miNA & 3900 & \miNA & \miNA \\
    2025 & BAT & NaS & \miNA & \miNA & \miNA & \miNA & \miNA & \miNA & \miNA & \miNA & \miNA & \miNA & \miNA & \miNA & \miNA & \miNA & \miNA & 1193 & \miNA & \miNA & \miNA & \miNA & \miNA \\
    2025 & ONW &  AG & \miNA & \miNA & \miNA & 1.72 & \miNA & \miNA & \miNA & \miNA & \miNA & \miNA & \miNA & 780.0 & 10.3 & 403.5 & \miNA & \miNA & \miNA & \miNA & \miNA & \miNA & \miNA \\
    2025 & ONW &  HTS-DD & \miNA & \miNA & \miNA & 1.72 & \miNA & \miNA & \miNA & \miNA & \miNA & \miNA & \miNA & 780.0 & 10.3 & 403.5 & \miNA & \miNA & \miNA & \miNA & \miNA & 1.00 & \miNA \\
    2025 & ONW &  SG-E-DD & \miNA & \miNA & \miNA & 5.16 & \miNA & \miNA & \miNA & \miNA & \miNA & \miNA & \miNA & 790.0 & 24.1 & 340.0 & \miNA & \miNA & \miNA & \miNA & \miNA & \miNA & \miNA \\
    2025 & ONW &  SG-PM-DD & \miNA & \miNA & \miNA & 13.7 & \miNA & \miNA & \miNA & \miNA & \miNA & \miNA & \miNA & 301.0 & 139 & 387.4 & \miNA & \miNA & \miNA & \miNA & 90.40 & \miNA & \miNA \\
    2025 & ONW &  SG-PM-HS & \miNA & \miNA & \miNA & 2.69 & \miNA & \miNA & \miNA & \miNA & \miNA & \miNA & \miNA & 304.3 & 26.2 & 440.0 & \miNA & \miNA & \miNA & \miNA & 90.40 & \miNA & \miNA \\
    2025 & ONW &  SG-PM-MS & \miNA & \miNA & \miNA & 3.51 & \miNA & \miNA & \miNA & \miNA & \miNA & \miNA & \miNA & 304.3 & 38.6 & 440.0 & \miNA & \miNA & \miNA & \miNA & 90.40 & \miNA & \miNA \\
    2025 & OFFW &  AG & \miNA & \miNA & \miNA & 1.72 & \miNA & \miNA & \miNA & \miNA & \miNA & \miNA & \miNA & 780.0 & 10.3 & 403.5 & \miNA & \miNA & \miNA & \miNA & \miNA & \miNA & \miNA \\
    2025 & OFFW &  HTS-DD & \miNA & \miNA & \miNA & 1.72 & \miNA & \miNA & \miNA & \miNA & \miNA & \miNA & \miNA & 780.0 & 10.3 & 403.5 & \miNA & \miNA & \miNA & \miNA & \miNA & 1.00 & \miNA \\
    2025 & OFFW &  SG-E-DD & \miNA & \miNA & \miNA & 5.16 & \miNA & \miNA & \miNA & \miNA & \miNA & \miNA & \miNA & 790.0 & 24.1 & 340.0 & \miNA & \miNA & \miNA & \miNA & \miNA & \miNA & \miNA \\
    2025 & OFFW &  SG-PM-DD & \miNA & \miNA & \miNA & 13.7 & \miNA & \miNA & \miNA & \miNA & \miNA & \miNA & \miNA & 301.0 & 141 & 240.0 & \miNA & \miNA & \miNA & \miNA & 90.40 & \miNA & \miNA \\
    2025 & OFFW &  SG-PM-HS & \miNA & \miNA & \miNA & 1.66 & \miNA & \miNA & \miNA & \miNA & \miNA & \miNA & \miNA & 304.3 & 20.0 & 440.0 & \miNA & \miNA & \miNA & \miNA & 90.40 & \miNA & \miNA \\
    2025 & OFFW &  SG-PM-MS & \miNA & \miNA & \miNA & 2.71 & \miNA & \miNA & \miNA & \miNA & \miNA & \miNA & \miNA & 304.3 & 38.6 & 440.0 & \miNA & \miNA & \miNA & \miNA & 90.40 & \miNA & \miNA \\
    2050 & PV & c-Si & 2.59 & \miNA & \miNA & \miNA & \miNA & \miNA & \miNA & \miNA & \miNA & \miNA & \miNA & \miNA & \miNA & 1.210 & \miNA & \miNA & \miNA & \miNA & \miNA & \miNA & \miNA \\
    2050 & PV & a-Si & \miNA & \miNA & \miNA & \miNA & \miNA & 14.8 & \miNA & \miNA & \miNA & \miNA & \miNA & \miNA & \miNA & 334.0 & \miNA & \miNA & \miNA & \miNA & \miNA & \miNA & \miNA \\
    2050 & PV & CIGS & \miNA & 0.00 & \miNA & \miNA & 1.66 & \miNA & 7.97 & \miNA & \miNA & \miNA & \miNA & \miNA & \miNA & \miNA & \miNA & \miNA & 11.7 & \miNA & \miNA & \miNA & \miNA \\
    2050 & PV & CdTe & \miNA & 19.4 & \miNA & \miNA & \miNA & \miNA & 0 & \miNA & \miNA & \miNA & \miNA & \miNA & \miNA & \miNA & \miNA & \miNA & \miNA & 21.9 & \miNA & \miNA & \miNA \\
    2050 & CSP & Parabolic Trough & 25.2 & \miNA & \miNA & \miNA & \miNA & \miNA & \miNA & \miNA & 74020 & \miNA & \miNA & 2000 & \miNA & 940.0 & \miNA & \miNA & \miNA & \miNA & 1.900 & \miNA & \miNA \\
    2050 & CSP & Fresnel Collector & 30.6 & \miNA & \miNA & \miNA & \miNA & \miNA & \miNA & \miNA & 85100 & \miNA & \miNA & 2000 & \miNA & 940.0 & \miNA & \miNA & \miNA & \miNA & 1.900 & \miNA & \miNA \\
    2050 & CSP & Solar Tower & 35.1 & \miNA & \miNA & \miNA & \miNA & \miNA & \miNA & \miNA & 58000 & \miNA & \miNA & 5700 & \miNA & 1800 & \miNA & \miNA & \miNA & \miNA & 1.700 & \miNA & \miNA \\
    2050 & EL & AEL & \miNA & \miNA & 6.70 & \miNA & \miNA & \miNA & \miNA & \miNA & 423.50 & \miNA & \miNA & \miNA & \miNA & 2922 & \miNA & \miNA & \miNA & \miNA & \miNA & \miNA & 108 \\
    2050 & EL & PEMEL & \miNA & \miNA & \miNA & \miNA & \miNA & \miNA & \miNA & 0.050 & \miNA & \miNA & \miNA & \miNA & \miNA & \miNA & 0.038 & \miNA & \miNA & \miNA & \miNA & \miNA & \miNA \\
    2050 & EL & HTEL\_SOEL-Y & \miNA & \miNA & \miNA & \miNA & \miNA & \miNA & \miNA & \miNA & \miNA & 76.8 & \miNA & \miNA & \miNA & 193.2 & \miNA & \miNA & \miNA & \miNA & \miNA & 0.47 & 71.9 \\
    2050 & BAT & NMC-111 & \miNA & \miNA & 410 & \miNA & \miNA & \miNA & \miNA & \miNA & \miNA & \miNA & 149 & 361.9 & \miNA & 414.0 & \miNA & \miNA & \miNA & \miNA & \miNA & \miNA & \miNA \\
    2050 & BAT & NMC-622 & \miNA & \miNA & 143 & \miNA & \miNA & \miNA & \miNA & \miNA & \miNA & \miNA & 132 & 132.5 & \miNA & 425.5 & \miNA & \miNA & \miNA & \miNA & \miNA & \miNA & \miNA \\
    2050 & BAT & NMC-811 & \miNA & \miNA & 92.3 & \miNA & \miNA & \miNA & \miNA & \miNA & \miNA & \miNA & 114 & 86.50 & \miNA & 739.2 & \miNA & \miNA & \miNA & \miNA & \miNA & \miNA & \miNA \\
    2050 & BAT & LFP & \miNA & \miNA & \miNA & \miNA & \miNA & \miNA & \miNA & \miNA & \miNA & \miNA & 92.4 & \miNA & \miNA & 64.50 & \miNA & \miNA & \miNA & \miNA & \miNA & \miNA & \miNA \\
    2050 & BAT & Redox-Flow & \miNA & \miNA & \miNA & \miNA & \miNA & \miNA & \miNA & \miNA & \miNA & \miNA & \miNA & \miNA & \miNA & \miNA & \miNA & \miNA & \miNA & \miNA & 3900 & \miNA & \miNA \\
    2050 & BAT & NaS & \miNA & \miNA & \miNA & \miNA & \miNA & \miNA & \miNA & \miNA & \miNA & \miNA & \miNA & \miNA & \miNA & \miNA & \miNA & 1193 & \miNA & \miNA & \miNA & \miNA & \miNA \\
    2050 & ONW &  AG & \miNA & \miNA & \miNA & 0.72 & \miNA & \miNA & \miNA & \miNA & \miNA & \miNA & \miNA & 780.0 & 4.32 & 403.5 & \miNA & \miNA & \miNA & \miNA & \miNA & \miNA & \miNA \\
    2050 & ONW &  HTS-DD & \miNA & \miNA & \miNA & 0.72 & \miNA & \miNA & \miNA & \miNA & \miNA & \miNA & \miNA & 780.0 & 4.32 & 403.5 & \miNA & \miNA & \miNA & \miNA & \miNA & 1.00 & \miNA \\
    2050 & ONW &  SG-E-DD & \miNA & \miNA & \miNA & 2.16 & \miNA & \miNA & \miNA & \miNA & \miNA & \miNA & \miNA & 790.0 & 10.1 & 340.0 & \miNA & \miNA & \miNA & \miNA & \miNA & \miNA & \miNA \\
    2050 & ONW &  SG-PM-DD & \miNA & \miNA & \miNA & 11.1 & \miNA & \miNA & \miNA & \miNA & \miNA & \miNA & \miNA & 301.0 & 113 & 387.4 & \miNA & \miNA & \miNA & \miNA & 90.40 & \miNA & \miNA \\
    2050 & ONW &  SG-PM-HS & \miNA & \miNA & \miNA & 1.30 & \miNA & \miNA & \miNA & \miNA & \miNA & \miNA & \miNA & 304.3 & 16.0 & 440.0 & \miNA & \miNA & \miNA & \miNA & 90.40 & \miNA & \miNA \\
    2050 & ONW &  SG-PM-MS & \miNA & \miNA & \miNA & 2.05 & \miNA & \miNA & \miNA & \miNA & \miNA & \miNA & \miNA & 304.3 & 32.0 & 440.0 & \miNA & \miNA & \miNA & \miNA & 90.40 & \miNA & \miNA \\
    2050 & OFFW &  AG & \miNA & \miNA & \miNA & 0.72 & \miNA & \miNA & \miNA & \miNA & \miNA & \miNA & \miNA & 780.0 & 4.32 & 403.5 & \miNA & \miNA & \miNA & \miNA & \miNA & \miNA & \miNA \\
    2050 & OFFW &  HTS-DD & \miNA & \miNA & \miNA & 0.72 & \miNA & \miNA & \miNA & \miNA & \miNA & \miNA & \miNA & 780.0 & 4.32 & 403.5 & \miNA & \miNA & \miNA & \miNA & \miNA & 1.00 & \miNA \\
    2050 & OFFW &  SG-E-DD & \miNA & \miNA & \miNA & 2.16 & \miNA & \miNA & \miNA & \miNA & \miNA & \miNA & \miNA & 790.0 & 10.1 & 340.0 & \miNA & \miNA & \miNA & \miNA & \miNA & \miNA & \miNA \\
    2050 & OFFW &  SG-PM-DD & \miNA & \miNA & \miNA & 11.1 & \miNA & \miNA & \miNA & \miNA & \miNA & \miNA & \miNA & 301.0 & 115 & 240.0 & \miNA & \miNA & \miNA & \miNA & 90.40 & \miNA & \miNA \\
    2050 & OFFW &  SG-PM-HS & \miNA & \miNA & \miNA & 1.30 & \miNA & \miNA & \miNA & \miNA & \miNA & \miNA & \miNA & 304.3 & 16.0 & 440.0 & \miNA & \miNA & \miNA & \miNA & 90.40 & \miNA & \miNA \\
    2050 & OFFW &  SG-PM-MS & \miNA & \miNA & \miNA & 2.05 & \miNA & \miNA & \miNA & \miNA & \miNA & \miNA & \miNA & 304.3 & 32.0 & 440.0 & \miNA & \miNA & \miNA & \miNA & 90.40 & \miNA & \miNA \\
    \bottomrule
    \end{tabular}
    \label{tab:materialIntensity}
\end{sidewaystable}